\documentclass[structabstract]{aa}

\usepackage{txfonts}
\usepackage{graphicx}
\usepackage{natbib}
\bibpunct{(}{)}{;}{a}{}{,}
\begin{document}

\title{Morphological effects on IR band profiles}
\subtitle{Experimental spectroscopic analysis with application to observed spectra of oxygen-rich AGB stars}

\author{A. Tamanai\inst{1} 
\and H. Mutschke \inst{1} 
\and J. Blum \inst{2} 
\and Th. Posch \inst{3} 
\and C. Koike \inst{4} 
\and J.W. Ferguson \inst{5}}

\institute{Astrophysical Institute and University Observatory,
Friedrich-Schiller-University Jena, Schillerg\"{a}\ss chen 3, 
D-07745 Jena, Germany
\and Institut f\"ur Geophysik und Extraterrestrische Physik,
Technische Universit\"{a}t Braunschweig, Mendelssohnstr.~3,
D-38106 Braunschweig, Germany
\and Institut f\"{u}r Astronomie, T\"{u}rkenschanzstrasse 17, 
A-1180 Wien, Austria
\and Department of Earth and Space Science, Graduate School of Science, 
Osaka University, 1-1 Machikaneyama, Toyonaka, Osaka 560-0043, Japan
\and Department of Physics, Wichita State University, Wichita KS67260-0032, 
USA}

      
\abstract{}
{To trace the source of the unique 13, 19.5, and 28~$\mu$m emission features in the spectra of oxygen-rich circumstellar shells around AGB stars, we have compared dust extinction spectra obtained by aerosol measurements.}
{We have measured the extinction spectra for 19 oxide powder samples of eight different types, such as Ti-compounds (TiO, TiO$_2$, Ti$_2$O$_3$, Ti$_3$O$_5$, Al$_2$TiO$_5$, CaTiO$_3$), $\alpha$-, $\gamma$-, $\chi$-$\delta$-$\kappa$-Al$_2$O$_3$, and MgAl$_2$O$_4$ in the infrared region (10~-~50~$\mu$m) paying special attention to the morphological (size, shape, and agglomeration) effects and the differences in crystal structure.} 
{Anatase (TiO$_2$) particles with rounded edges are the possible 13, 19.5 and 28~$\mu$m band carriers as the main contributor in the spectra of AGB stars, and spherically shaped nano-sized spinel and Al$_2$TiO$_5$ dust grains are possibly associated with the anatase, enhancing the prominence of the 13~$\mu$m feature and providing additional features at 28~$\mu$m. The extinction data sets obtained by the aerosol and CsI pellet measurements have been made available for public use at {\it http://elbe.astro.uni-jena.de}.}
{}


\maketitle

\section {Introduction}
Dust grains, which are composed mostly of micron-sized 
solid particles, are crucial players for the 
initial stages of star and planet formation.
A significant dust production source is in the outflow 
of asymptotic giant branch (AGB) stars 
which lose almost 90~\% of their total mass during mass loss at the end of their life time
\citep[e.g.][]{sed94, GS99}. 
Dust formation can only occur if the temperature is low enough and 
the density of condensable molecules is high enough for condensates to be 
stable against evaporation. 
Regions with low temperatures usually occur only at a significant 
distance above the photosphere with densities that are orders of magnitude 
lower than in the photosphere of stars \citep{HO04}.  
Notably, observed emission spectra of AGB stars provide 
information on the physical and chemical properties of dust grains present. 
Nevertheless, the chemical and mineralogical composition 
of dust grains in AGB stars are not yet well understood. 
Depending on temperature and pressure conditions, 
high temperature condensates (HTCs) 
such as corundum, spinel and perovskite \citep[e.g.][]{GL74, SH90}
are most important because they contribute significant 
opacity when no other grains are present, even though 
they are not as abundant as the silicate or iron grains which form at lower temperatures. 
Once the more abundant grains begin to condense, these 
less abundant grains lose their significance. Although 
they are significant only over a narrow range of temperatures, 
it is an important transition region between the molecular 
regime and the dust regime \citep{Feg05}. \\
\indent
The strong and unique 13 and 19.5~$\mu$m emission features 
detected from oxygen-rich (O-rich) circumstellar 
envelopes around evolved stars may conceivably be caused by HTCs. \\
\indent
A crystalline form of Al$_2$O$_3$ has been considered as a carrier of the 13~$\mu$m feature \citep[e.g.][]{Ona89,Gla95,Beg97,Slo03,Str04,De06}. \citet{Beg97} demonstrated that amorphous Al$_2$O$_3$, which was fabricated by a sol-gel technique, is not able to explain the 13~$\mu$m feature of circumstellar O-rich envelopes. However, they suggested that crystalline $\alpha$-Al$_2$O$_3$ may be the 13~$\mu$m band carrier if the particle shape is taken into consideration. Silicate dust grains might be closely related to the 13~$\mu$m band as well. \citet{Slo03} investigated the correlated dust features at 13, 20, and 28~$\mu$m of O-rich circumstellar shells and suggested that crystalline Al$_2$O$_3$ is the 13~$\mu$m feature carrier rather than spinel, and that the 20 and 28~$\mu$m features are contributed by silicates. 
\citet{Str04} also proposed that crystalline corundum seems to be a promising candidate for the 13~$\mu$m feature due to the structure and composition of two formes presolar Al$_2$O$_3$ grains which were discovered in the Tieschitz ordinary chondrite. These structural and compositional differences are directly linked to the condensation sequence of dust grains and reflect the observed spectra. There is absence of the 22~$\mu$m feature from crystalline corundum in observed spectra as well. \citet{De06} studied the IR spectrum features at 13 and 21~$\mu$m of AGB stars using the one-dimensional radiative transfer code DUSTY. 
The target species were corundum, spinel, silicate, and 
amorphous alumina, and included the effects of grain shape and relative 
abundances of mixing samples. 
They found that corundum fit well as the 13~$\mu$m feature carrier 
rather than spinel if the grains have a spherical shape. \\
\indent
However, \citet{Po99} have shown that neither corundum nor rutile could account for the 13~$\mu$m emission band because the peak position of corundum was located at slightly shorter wavelength (12.7~$\mu$m). Only spherical spinel corresponded closely to the 13~$\mu$m emission band (12.95~$\mu$m) in their theoretical calculations. 
\citet{Fab01} have examined experimentally the spectra of synthesized spinel by taking different Al/Mg-ratios into account. They found out that the peak position of the near-stoichiometric synthetic spinel (e.g.~Mg$_{1.01}$Al$_{1.99}$O$_4$ and Mg$_{0.94}$Al$_{2.04}$O$_{4}$) fit well at 13~$\mu$m compared with the spectra of O-rich circumstellar shells. \\
\indent
\citet{Sp00}, on the other hand, proposed silicon dioxide (SiO$_2$) or polymerized silicates as a 13~$\mu$m feature carrier candidate. They suggested that the 13~$\mu$m feature is contributed by the silicates in a similar way to \citet{Beg97}. \citet{Sp00} exhibited two polytypes of SiO$_2$ which produced a comparatively clear 13~$\mu$m band and also pointed out that a detectable amount of SiO$_2$ formation is feasible considering the atomic abundances. \\
\indent
Heterogeneous dust grains have been considered for the identification of the 13~$\mu$m feature as well. \citet{KS97} proposed core-mantle grains which were composed of an $\alpha$-Al$_2$O$_3$ core surrounded by a silicate mantle as a possible candidate for the 13~$\mu$m carrier. \citet{Po99} examined the 13~$\mu$m band profile by making use of the  Maxwell-Garnett theory. When the volume fraction of corundum was 0.85 (corundum core 85~\% and amorphous olivine mantle 15~\%), the peak position in the spectrum was at 12.95~$\mu$m in wavelength. \\
\indent
Based upon a model calculation of O-rich circumstellar dust shells around pulsating AGB stars, \citet{Je99} stated that TiO$_2$ would be the most promising candidate as the nucleation seed for further heterogeneous grain growth instead of Al$_2$O$_3$. Until now, the presence of TiO$_2$ dust grains in spectra of AGB stars has not been confirmed; however, they were identified in meteorites such as carbonaceous chondrites \citep{Gre96,Gre98} and the Krymka unequilibrated ordinary chondrites \citep{Nit05}. 
Presolar titanium oxide in unequilibrated ordinary chondrite meteorites has been reported \citep{Nit08}. 
\citet{Gre96} suggested that 
these oxides did not undergo further chemical reactions 
with remaining nebular gas during the formation of different oxide 
or silicate dust grains. 
According to elemental abundances in meteorites \citep{Ca73}, magnesium (Mg), silicon (Si), and iron (Fe) are more abundant than aluminum (Al) and calcium (Ca) by a factor of 12. 
Titanium (Ti) is a much less abundant metal; it is a 
factor of 400 times less abundant than Si. Thus Mg-, 
Si-, Fe-, Al-, and Ca-compounds surpass Ti-compounds 
in quantity. \\
\indent
Experimentally measured spectra and theoretical calculations which assume simple geometrical models such as spherical \citep{mie08} or ellipsoidal \citep{bh83} grain shapes have been generally applied to compare models with observed spectra. It has recently become possible to utilize specific methods for particularly inhomogeneous structures and arbitrarily shaped particles for absorption, scattering and extinction calculations \citep[e.g.][]{pp73,dr88,mi90,mm96,min05}. However, predictions based on these calculations are uncertain since, in reality, grain shape might be tremendously irregular and complex. \\
\indent
Regarding experimental approaches, absorption spectra have been measured by the KBr (potassium bromide) pellet technique for the mid-IR region in most cases \citep[e.g.][]{do78,Ko81,Oro91,Po99,Chi02}. A controversial point of this technique is that the band profile is substantially changed by the influence of its electromagnetic polarization since a solid sample is embedded in a medium (KBr) \citep{Fab01}. \\
\indent
In this paper, we have experimentally investigated the extinction spectra of $\alpha$-, $\gamma$-, $\chi$-$\delta$-$\kappa$-Al$_2$O$_3$, MgAl$_2$O$_4$, and Ti-compounds (TiO, TiO$_2$, Ti$_2$O$_3$, Ti$_3$O$_5$, Al$_2$TiO$_5$, CaTiO$_3$) in the mid-IR region (10-50~$\mu$m) to clarify the possible 13, 19.5 and 28~$\mu$m band carriers with special consideration of morphological effects by the use of the aerosol technique \citep[e.g.][]{Hi99} so as to avoid the influence of electromagnetic interaction with solid embedding media \citep{ta06a,ta06b}. The aim is to obtain 
band profiles without the medium effect of these HTCs for a direct 
comparison to observed spectra and to identify the possible candidates. 
\section{Experiment}
\subsection{Classical vs. aerosol techniques}
The pellet technique is the classic technique, where a solid sample is mixed with potassium bromide (KBr), cesium iodide (CsI), or polyethylene (PE) powder that have high transmission through certain IR wavelength ranges, and the mixture is pressed with a 10~Ton load in order to make a 0.55~mm thick 
(1.2~mm for PE) and 13~mm diameter pellet for spectroscopic analysis \citep[e.g.][]{KH87,Ja94,Beg97,Chi02}. The advantages of the pellet technique are low cost, low sample consumption, longevity of the pellets in a desiccator and the exact amount of a measured sample is known. 
On the other hand, the main disadvantage is that there is the possibility of environmental effects due to the electromagnetic polarization of the embedding medium \citep[e.g.][]{Pa98,HM00,Sp00,Cl03}. 
\citet{ta06a,ta06b} introduced the aerosol technique 
into dust grain investigation and demonstrated that 
the strong absorption peaks at approximately 9.8 and 
11~$\mu$m obtained by aerosol measurements for an 
olivine-type crystalline powder are shifted to shorter 
wavelengths by amounts of up to 0.24 $\mu$m compared 
with spectra obtained by the KBr pellet 
measurements. 
Conversely, weak features are not affected much by  
the KBr medium effect \citep[see more details in][]{ta06b}. \\
\begin{figure*}
\begin{center}
\includegraphics[scale=0.5]{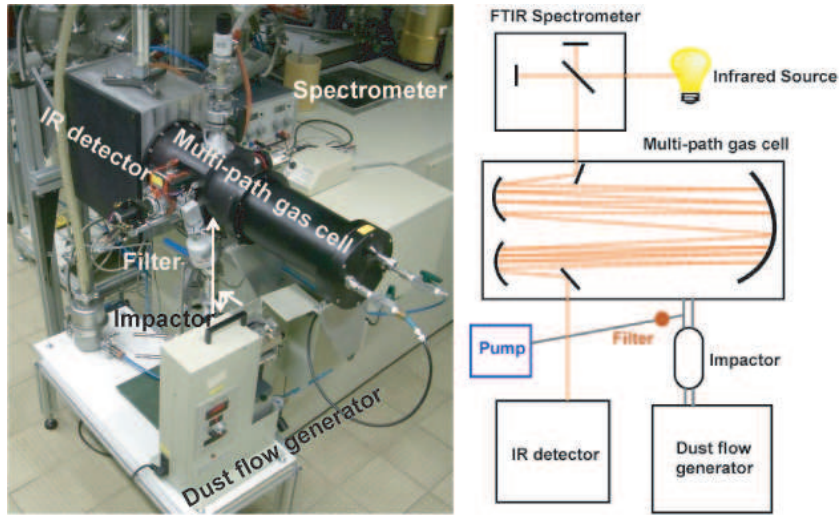}
\caption[\small Experimental apparatus for aerosol extinction measurements]{Experimental apparatus for aerosol extinction measurements. A photo (left) and a schematic diagram (right) of experimental device setup.}
\end{center}
\end{figure*}
\indent
Fig.~1 shows the setup of the apparatus for the aerosol measurement. We have utilized a dust flow generator (Palas RBG1000) to disperse a powdered sample in a nitrogen (N$_2$) gas stream. The dense aerosol is carried to a two-stage impactor which separates the large particles from the small ones (d$_{\rm avg}$$\approx$~2-3~$\mu$m). 
The small-grained aerosol are concentrated by the impactor so that a concentration of 10$^6$ particles per cubic centimeter finally arrives at a White-type long-path infrared cell (MARS-8L/20V, Gemini Scientific Instr.) 
which has an 18~m path length by taking advantage of multiple reflections  
between two gold mirrors mounted on both sides in order to increase the sensitivity. 
Since a Fourier Transformation Infrared Spectrometer (FTIR, Bruker 113v) with a DTGS detector with CsI windows is fixed to the cell, it is possible to measure the extinction spectrum of the suspended dust particles in N$_2$ gas. As a consequence, we avoid any environmental effect during the aerosol measurement, and the measurement conditions are closer to a vacuum with regard to the dielectric function of the medium (N$_2$: $\epsilon$$\approx$1.0) compared to KBr ($\epsilon$=2.3). Moreover there is a risk of deforming the powdered sample structure by the high pressure required for the KBr pellet technique, which is avoided when the aerosol technique is utilized for the measurement. \\
\indent
To make the environmental effect clearer, we have used CsI powder to create pellets as well. CsI is particularly transparent between 2 and 50~$\mu$m wavelengths, and has a greater optical constant than that of KBr (CsI 1.74 \& KBr 1.52). We have extended the wavelength range up to 50~$\mu$m for both aerosol and CsI pellet measurements.
The aerosol particles are extracted on a polyester-membrane 
filter which is mounted externally between the outlet of the impactor and 
the cell. The morphological properties of extracted particles 
are investigated with a scanning electron microscope (SEM).
\subsection{Samples}
We have investigated 19 powdered crystalline samples of eight different combinations of composition and crystal structure. 
\begin{table*}
\normalsize
\caption{Properties of the samples}
\centering
\small
\begin{tabular}{c c c c c c c }
\hline
Chem. formula & Mineralogical name & Product Info. & Abbr. & Size ($\mu$m) & Shape & Sect. \\
\hline
TiO     & Titanium monoxide & Aldrich           & TiO & 0.1-2.0         & irr. w/ sharp edges           & A.1. \\
\hline
$\alpha$-TiO$_2$  & Rutile & Aldrich            & CR1 & 0.1-0.5         & irr. w/ round. edges          & A.2. \\
$\alpha$-TiO$_2$  & Rutile & Alfa Aesar         & CR2 & 0.1-0.5         & irr. w/ sharp \& round. edges & A.2. \\
$\alpha$-TiO$_2$  & Rutile & Tayca 150W         & CR3 & 0.01-0.08       & thin \& long                  & A.2.\\
$\alpha$-TiO$_2$ & Rutile  & Chitan Inc. No.101 & CR4 & 0.01-0.1        & sph. \& ellip.                & A.2.\\
$\beta$-TiO$_2 $ & Anatase & Alfa Aesar         & CA1 & 0.1-0.6         & round. edges                  & A.2. \\
$\beta$-TiO$_2 $ & Anatase & C.P.               & CA2 & 0.1-0.2         & round. \& squ. \& ellip.      & A.2. \\
$\beta$-TiO$_2 $ & Anatase & Tayca 600          & CA3 & 0.01-0.05       & round.                        & A.2. \\
$\beta$-TiO$_2 $ & Anatase & Chitan Inc. No.9   & CA4 & $\approx$~0.05  & round. \& long-narrow         & A.2. \\
\hline
Ti$_2$O$_3$ & Dititanium trioxide & Alfa Aesar & Ti$_2$O$_3$ & 0.2-1.5  & irr. w/ sharp edges         & A.3. \\
\hline
Ti$_3$O$_5$ & Trititanium pentoxide & Alfa Aesar & Ti$_3$O$_5$ & 0.1-1.0 & irr. w/ sharp edges        & A.4. \\
\hline
CaTiO$_3$   & Perovskite    & Alfa Aesar    & CaTiO$_3$ & 0.03-0.5      & irr. w/ sharp edges         & A.5. \\
\hline
Al$_2$TiO$_5$ & Tialite & Alfa Aesar   & Al$_2$TiO$_5$ & 0.05-0.6       & irr. w/ very sharp edges    & A.6. \\
\hline
$\alpha$-Al$_2$O$_3$  & Corundum & Alfa Aesar & CAC1 & 0.3-0.5        & irr. w/ round. edges          & A.7. \\
$\alpha$-Al$_2$O$_3$  & Corundum & Glaschemie Jena & CAC2 & 0.06-1.0  & irr. w/ sharp edges           & A.7. \\
$\gamma$-Al$_2$O$_3$ & & Alfa Aesar & CG & 0.1-0.3                    & squ. w/ round. edges          & A.7. \\
$\chi$-$\delta$-$\kappa$-Al$_2$O$_3$ & & C.P. & CCDK & 0.05-1.0       & irr. \& tabular               & A.7. \\
\hline
MgAl$_2$O$_4$ & Spinel & Alfa Aesar & CSp1 & 0.04-0.5                 & irr. w/ sharp \& round. edges & A.8. \\
MgAl$_2$O$_4$ & Spinel & Aldrich    & CSp2 & $\approx$ 0.05           & sph.                          & A.8. \\
\hline
\end{tabular}
\begin{center} 
\tiny "C.P." $\rightarrow$ commercial product; "irr." $\rightarrow$ irregular; "round." $\rightarrow$ roundish; "squ." $\rightarrow$ square; "ellip." $\rightarrow$ ellipsoidal; "sph." $\rightarrow$ spherical
\end{center}
\end{table*}
In particular, for TiO$_2$ and Al$_2$O$_3$, special attention is paid to the difference in the crystal structures, and MgAl$_2$O$_4$ and TiO$_2$ are utilized for morphological examination. Table~1 gives a list of the investigated samples including their properties. All the samples are commercial products. Since our aerosol apparatus is highly effective only for particle sizes of less than about 1~$\mu$m, Al$_2$TiO$_5$ (original size $\approx$ 149~$\mu$m) and Ti$_3$O$_5$ (original size 0.1-0.3~mm) were milled by a ball mill process (Si$_3$N$_4$ balls for 30~min.) to obtain  particle sizes less than 1~$\mu$m. In addition, a size fraction $<$ 1~$\mu$m in diameter for $\chi$-$\delta$-$\kappa$-Al$_2$O$_3$ is concentrated by sedimentation in a solvent (acetone). \\
\indent
Detailed information of each oxide is given in Appendix A. \\
\subsection{Equation of state}
The equation of state (EOS) of the PHOENIX stellar atmosphere code 
can calculate the chemical equilibrium (gas and dust in equilibrium) 
of 40 elements, including the ionization stages, with hundreds of 
molecular, liquid, and solid species 
\citep[see details in][]{Al01,Feg05}. 
Although the EOS does not provide information 
on the chemical pathways, it is possible to obtain number fractions 
of dust species at specific temperature and pressure points.
We use the PHOENIX EOS for our spectroscopic investigation since it 
can predict the abundance patterns of various dust species in 
the outflow of AGB stars.
Fig.~2 shows the abundances of condensates as a function of temperature 
for a single gas pressure, 10$^{4}$~dyne/cm$^2$. 
The first dust species to condense are four different crystal structures 
of Al$_2$O$_3$ at approximately 1800~K. 
The most abundant one is $\alpha$-Al$_2$O$_3$ 
followed by $\gamma$-, $\kappa$-, $\delta$-Al$_2$O$_3$. 
Then, perovskite appears at around 1700~K. 
Solid Ti-compounds such as Ti$_2$O$_3$ and TiO$_2$ come 
out of the gas phase below 1600~K. 
TiO$_2$ appears below 1000~K. 
However, Ti-compounds are less abundant than 
silicates by a factor of 1000 because Ti is a factor 
of 400 less abundant metal than Si \citep{Ca73}.
The most abundant species above 1500~K is spinel which 
appears at about 1600~K in this calculation. 
Mg-rich silicate grains dominate below 1500~K (Note: We 
plotted only 12 species that have been investigated in 
our experiments here).\\
\begin{figure}
\begin{center}
\includegraphics[scale=0.3]{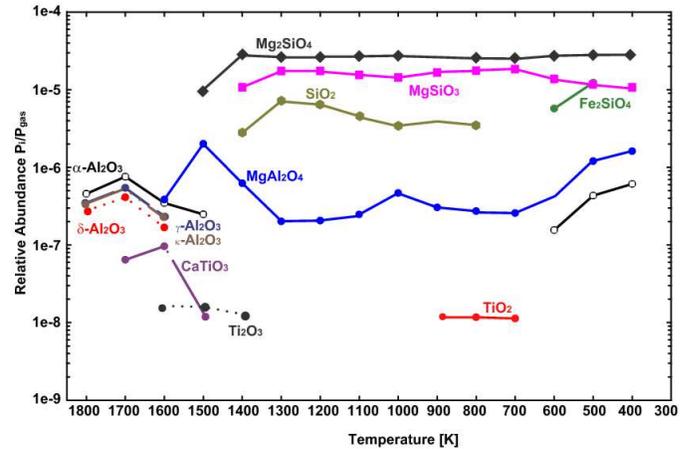}
\caption[\small EOS]{Relative abundances of dust grain species for a single gas pressure (10$^{4}$~dyne/cm$^{2}$) as a function of temperature.}
\end{center}
\end{figure}
\indent
We concentrate on those HTCs (Al$_2$O$_3$, spinel, and Ti-compounds) 
for experimental spectroscopic analysis (Note: We call HTCs in this 
paper those species for which dust grains condense out of the gas 
phase above 1500~K in this EOS calculation, plus TiO$_2$). 
\section{Spectroscopic results}
\subsection{Aerosol versus CsI pellet spectra}
The aerosol spectra reveal substantially different band 
profiles in comparison with the CsI pellet measurements. 
\citet{ta06b} demonstrated the disparity between aerosol 
and KBr pellet spectra of various silicate samples. 
Fig.~3 shows the three extinction spectra of rutile (CR1) 
up to 25~$\mu$m obtained by aerosol, CsI, and KBr pellet 
measurements. 
The strongest peak at 13.53~$\mu$m of the aerosol spectrum 
shifts to 15.61~$\mu$m in wavelength with the CsI pellet 
measurement ($\Delta$$\lambda$=2.08~$\mu$m). The shift is 
larger than in the case of the KBr spectrum which is explained 
by the relationship between the dielectric functions of a sample 
($\epsilon$) and the embedding medium ($\epsilon$$_{m}$) \citep[see][]{bh83}. 
As the value of $\epsilon$$_{m}$ increases (N$_2$ 1.0; KBr 2.3; CsI 3.0), 
the peak positions are significantly shifted to longer wavelengths 
by the influence of its electromagnetic polarization. \\
\begin{figure}
\begin{center}
\includegraphics[scale=0.45]{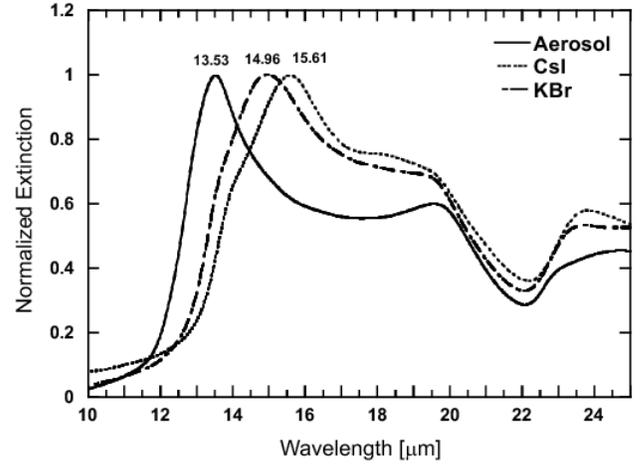}
\caption[\small Normalized extinction vs. wavelength of TiO$_2$ (CR1) spectra]{Normalized extinction vs. wavelength of TiO$_2$ (CR1) spectra obtained by aerosol (solid line), CsI (dotted line), and KBr pellet measurements (dash-dotted line).}
\end{center}
\end{figure}
\indent
Fig.~4 shows the extinction spectra in the wavelength range 
between 10 and 50~$\mu$m for all 19 samples which are plotted as 
normalized extinction (except TiO).
The extinction spectra are normalized because aerosol measurements 
are not quantitative. 
TiO has a featureless spectrum in both the
aerosol and CsI measurements because it is dominated by free 
charge carrier absorption.
The rutile spectra are characterized by a strong, often double-peaked,
absorption band between 12 and 22 $\mu$m (as already shown in Fig. 3), 
a second absorption band between 22 and 27 $\mu$m, and a third very broad one between 
27 and 50 $\mu$m, which is much more centered at the short-wavelength 
side for the CsI spectra. For one sample (CR3), these three bands are 
merged into a very broad complex. 
The strong difference between these spectra and those calculated by \citet{Po99} and
\citet{Po03} for spherical grains will be discussed in the next Section.
Compared to rutile (TiO$_2$), anatase (TiO$_2$) lacks the second of the bands and the first
one instead extends to somewhat longer wavelengths ($\approx$~27 $\mu$m). 
Also, the 12-27 $\mu$m band in two cases (CA3 \& CA4) shows strongly 
pronounced differences between the aerosol and the CsI spectra. 
While in the first case it is double-peaked, in the second it has only 
a single-peaked profile with the peak at 16-18 $\mu$m.
On the other hand, the long-wavelength band has not been observed to be so
strongly broadened beyond 40 $\mu$m as seen for the rutile samples CR1
and CR4.\\
\indent
The calculations for spherical particles presented by \citet{Po03}
produce a double band peaking at 13 and 15 $\mu$m and a sharp 27.5 $\mu$m band.
The CA1 and CA2 aerosol spectra come relatively close to this prediction,
although the profiles are much broader (see next section).
Ti$_2$O$_3$ has a double-peaked band between 17 and 21 $\mu$m and only
weak features longward of it. The aerosol-measured spectrum reproduces 
very well the calculated spectrum for spherical particles \citep{Po03}. 
The (normalized) spectrum of CsI embedded particles shows an enhanced 
far-infrared extinction, which is probably due to conductivity and certainly is 
enhanced by agglomeration within the pellet (see also the TiO spectrum). 
The Ti$_3$O$_5$ spectrum does not show this effect, but is complicated in 
terms of the appearance of many peaks. 
These peaks do not group into clear complexes, unlike the simpler titanium oxides.\\
\indent
Perovskite again has a divided spectrum with a relatively narrow
short-wavelength band, the peak of which shifts strongly with the embedding medium. 
Even for the aerosol spectrum, the peak is at longer wavelengths compared to the 
calculated spectrum for spherical grains presented by \citet{Po03}, 
whereas the longer-wavelength bands reproduce the calculated band positions quite well. 
The aluminum titanate spectrum lacks this division and is in fact reminiscent of 
the $\gamma$-Al$_2$O$_3$ spectrum with some additional smaller structures,
but not as many as the $\chi$-$\delta$-$\kappa$-Al$_2$O$_3$. \\
\indent
\citet{Ko95} performed extinction efficiency 
measurements of two different $\gamma$-Al$_2$O$_3$ samples 
(commercial aerosil particles and a combustion product). 
Both samples showed a double peak at approximately 12.4 and 
13.9~$\mu$m. 
Similarly, \citet{Ku05} carried out transmittance measurements 
of commercial $\gamma$-Al$_2$O$_3$ particles, and they detected 
a double peak at about 12.5 and 13.5~$\mu$m in wavelengths as well. 
These results fit well with our measurements (aerosol 12.3-13.1~$\mu$m 
\& CsI 12.7-13.6~$\mu$m). 
Moreover, the aerosil spectrum from Koike et al.~(1995) exhibited 
also a small shoulder around 17-18~$\mu$m. 
The small shoulder is apparent in our $\gamma$-Al$_2$O$_3$ spectra 
obtained by both the aerosol (15.7~$\mu$m) and CsI pellet (16.6~$\mu$m) 
measurements. 
The slight differences in peak positions come from mainly two factors. 
One is the medium effect since all observations utilized different 
substances as a medium. 
\citet{Ko95} and \citet{Ku05} used KBr powder whereas we applied N$_2$ 
gas for the aerosol measurements and CsI for the pellet measurements. 
Another factor is the morphological effect. 
Our $\gamma$-Al$_2$O$_3$ particles are square shape with rounded edges 
and 0.1-0.3~$\mu$m in size while the aerosil particles \citep{Ko95} were 
disk shape with a mean diameter of 0.02~$\mu$m. \\
\indent
Corundum shows two strong band complexes at 11-18 and 19-23
$\mu$m plus a small band at 26~$\mu$m. 
Both complexes have clear substructures, although the structure 
strongly depends on the sample (see below). 
The spectra do not show dominant sharp peaks at 13 $\mu$m or 
shortward of 13 $\mu$m, which would be expected for spherical grains according 
to Mie calculations \citep{Beg97,Po99}.\\
\indent
Another important point for Al$_2$O$_3$ is the state of its crystal structure. 
Though it is obvious to confirm the morphological differences among these 
Al$_2$O$_3$ samples via the SEM and TEM images, these band profiles are 
significantly affected by differences in the crystal structures, 
in this case more than by the morphological effects. \\
\indent
\citet{Ku05} produced nano-sized spherical 
$\delta$-Al$_2$O$_3$ particles by the gas evaporation method 
and measured the transmittance spectra by making use of the 
KBr pellet technique. 
They detected 15 absorption peaks in the wavelength range 
between 10 and 19~$\mu$m. 
$\delta$-Al$_2$O$_3$ showed many sharp peaks in an exceptionally 
broad absorption region (10-20~$\mu$m). 
A similar result is obtained in our 
$\chi$-$\delta$-$\kappa$-Al$_2$O$_3$ sample investigation up 
to 20~$\mu$m, and more peaks are observed in both the aerosol 
and CsI pellet spectra up to 46~$\mu$m.\\
\indent
The relation between cubic alumina and (non-stoichiometric) spinel
spectra has already been discussed by \citet{Fab01} based on 
reflection spectra. It is interesting to note that the aerosol measurement 
of $\gamma$-alumina shows a dominance of the 12.5 $\mu$m peak in the spectrum and 
a shift of this feature to shorter wavelengths compared to the CsI measurement. 
The spectrum comes rather close to that calculated by \citet{Beg97} 
for a shape distribution from the optical constants presented by \citet{Ko95}, 
although an additional shoulder at 15.7 $\mu$m is seen. This shoulder becomes a 
strong band in the CsI spectrum.\\
\indent
The spinel spectra show two main bands at about 13-14 $\mu$m and 17-19 $\mu$m
depending on sample and embedding. 
The band at 32 $\mu$m is also clearly seen in
both the aerosol and CsI spectra. 
Based on calculations for spherical grain shapes,
\citet{Fab01} proposed that synthetic and also annealed natural spinel
particles would reproduce the 13 $\mu$m band of AGB stars. 
Our samples CSp1 and CSp2
both are synthetic material. The CSp2 spectrum measured in aerosol peaks at 13.26 $\mu$m,
but when measured in CsI, the peak is shifted to more than 14 $\mu$m, which is similar to
the position reported by \citet{Po99} for spinel particles in KBr. 
The CSp1 spectra peak at more than 14 $\mu$m in all cases, which will be discussed in the next
section.\\
\indent
In general, there is a clear trend that the aerosol spectra have peaks
at significantly shorter wavelengths than the CsI spectra. 
In particular, this is always true for the shortest-wavelength peak. 
However, as already mentioned for anatase, in most cases
the matrix effect cannot be reduced to a simple peak shift. Specifically 
for strong bands, the band profiles measured in aerosol sometimes show 
a broadening to longer wavelengths as well, sometimes resulting in a 
secondary peak and a near-rectangular total profile. 
These effects are seen to a much lesser extent in the CsI spectra. 
This may partially also be a consequence of 
differences in the particle morphology, which may have been caused by\\
(a) using different dispersion methods;\\ 
(b) particles may transform during the grinding procedure; \\
(c) powdered sample structure deformation caused by the high pressurization required for the CsI pellet technique. \\
A large difference in band profiles between the aerosol and CsI pellet measurements is possibly produced more by spherical and ellipsoidal shapes than irregular ones.
\subsection{Morphological effects}
One of the advantages of using the aerosol technique 
in the investigation of dust is that it is possible to inspect 
the particle morphology. 
We set a polyester-membrane filter between the outlet of the 
impactor and the cell (Fig.~1) and extract the incoming aerosol 
particles. 
These captured aerosol particles are analyzed by a SEM. 
The SEM images are shown in Fig.~4 (right). 
A TEM has also been utilized to understand the individual particle 
shape as well (Fig.~4 middle images). \\
\begin{figure*}
\begin{center}
\includegraphics[scale=0.8]{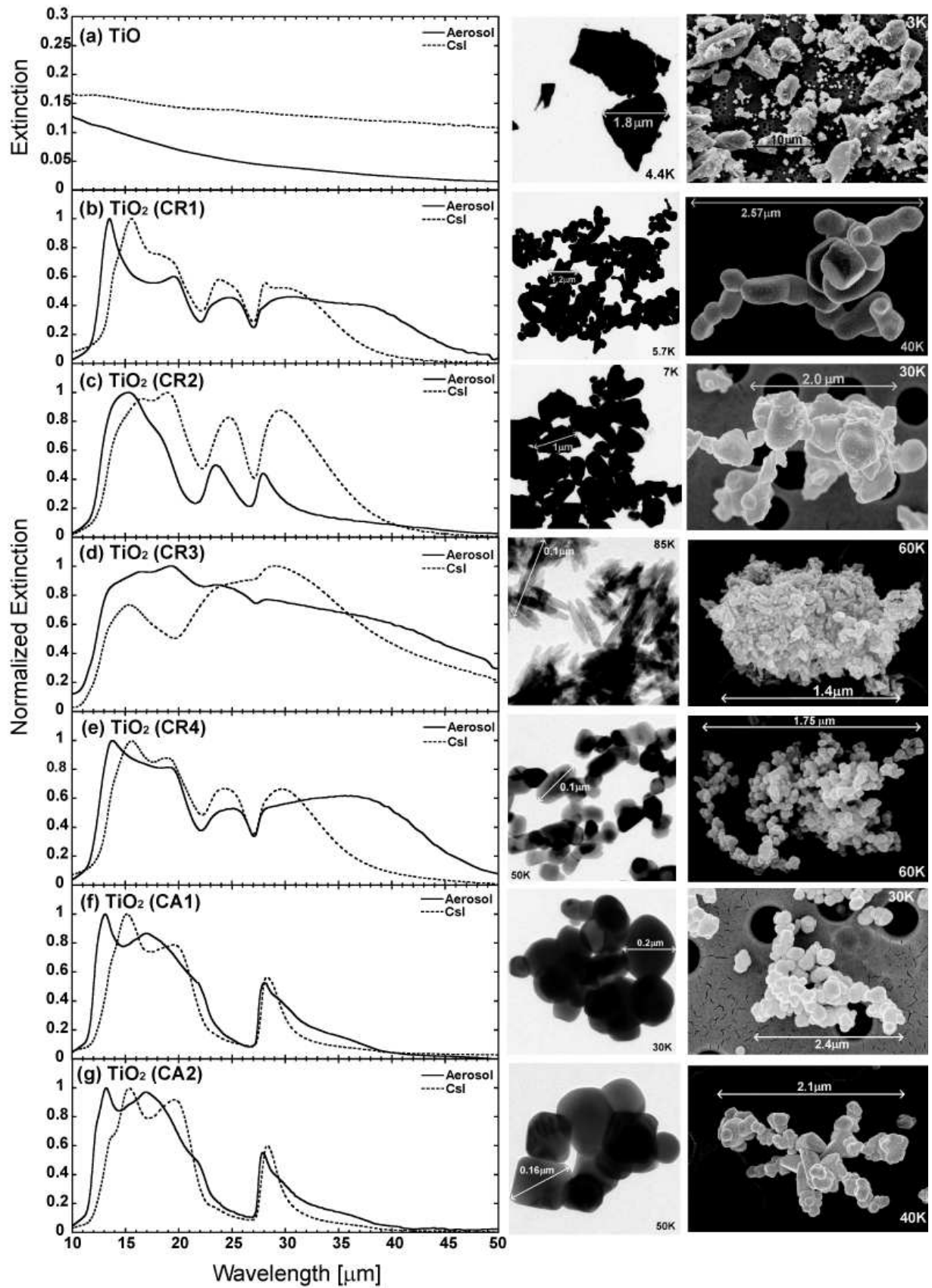}
\caption[\small Normalized extinction vs. wavelength of 19 HTCs]{Left: Normalized extinction vs. wavelength of 
(a) TiO; (b) TiO$_2$ (CR1); (c) TiO$_2$ (CR2); (d) TiO$_2$ (CR3); (e) TiO$_2$ (CR4); (f) TiO$_2$ (CA1); (g) TiO$_2$ (CA2) spectra obtained by aerosol and CsI pellet measurements. Center: TEM images (original powder). Right: SEM images (aerosol particles). Note that only the TiO spectrum is not normalized.}
\end{center}
\end{figure*}
%
\begin{figure*}
\begin{center}
\includegraphics[scale=0.8]{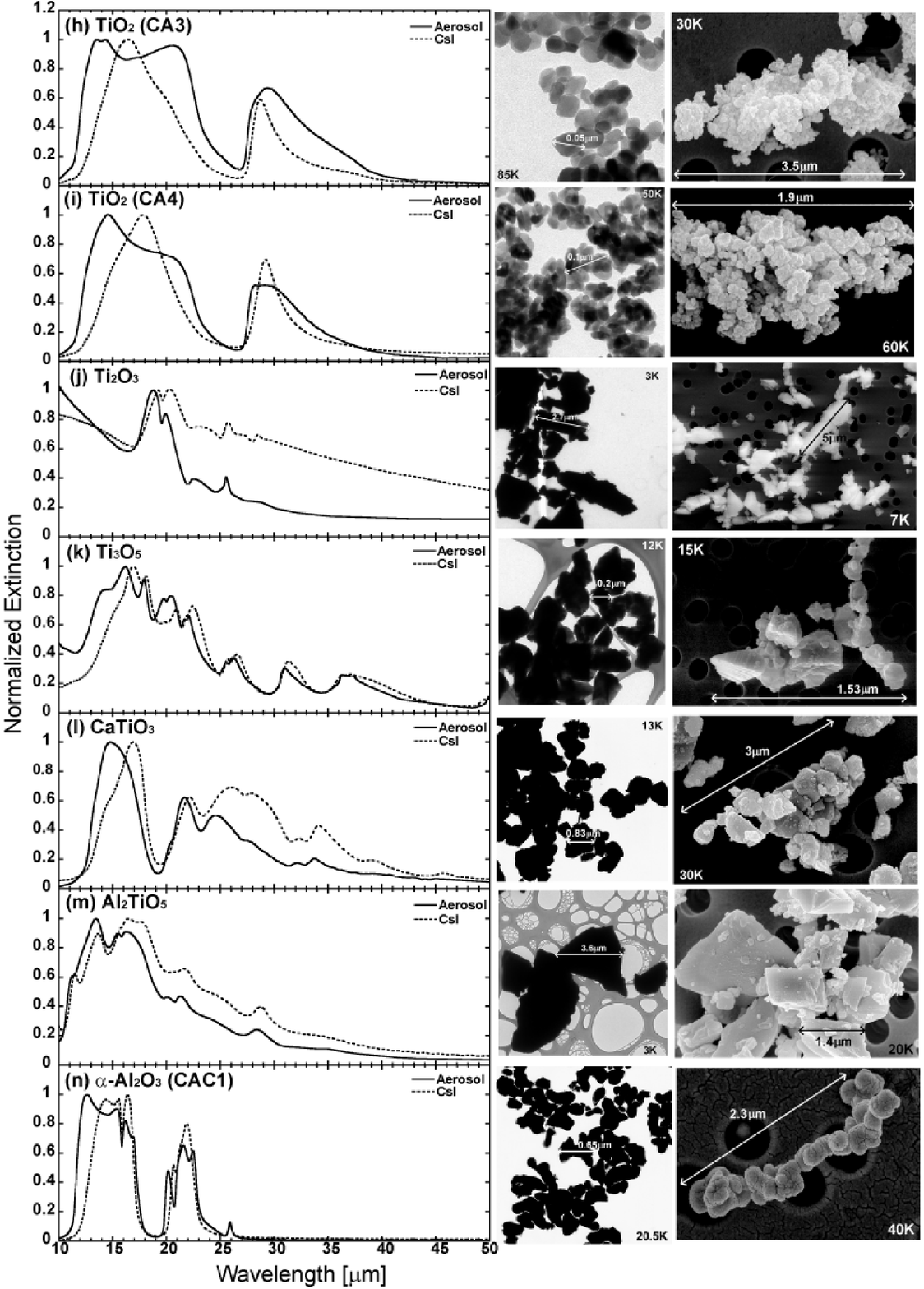}
\addtocounter{figure}{-1}
\caption[\small Normalized extinction vs. wavelength of 19 HTCs]{({\it continued}) Left: Normalized extinction vs. wavelength of (h) TiO$_2$ (CA3); (i) TiO$_2$ (CA4); (j) Ti$_2$O$_3$; (k) Ti$_3$O$_5$; (l) CaTiO$_3$; (m) Al$_2$TiO$_5$; (n) $\alpha$-Al$_2$O$_3$ (CAC1) spectra obtained by aerosol and CsI pellet measurements. Center: TEM images (original powder). Right: SEM images (aerosol particles).}
\end{center}
\end{figure*}
%
\begin{figure*}
\begin{center}
\includegraphics[scale=0.8]{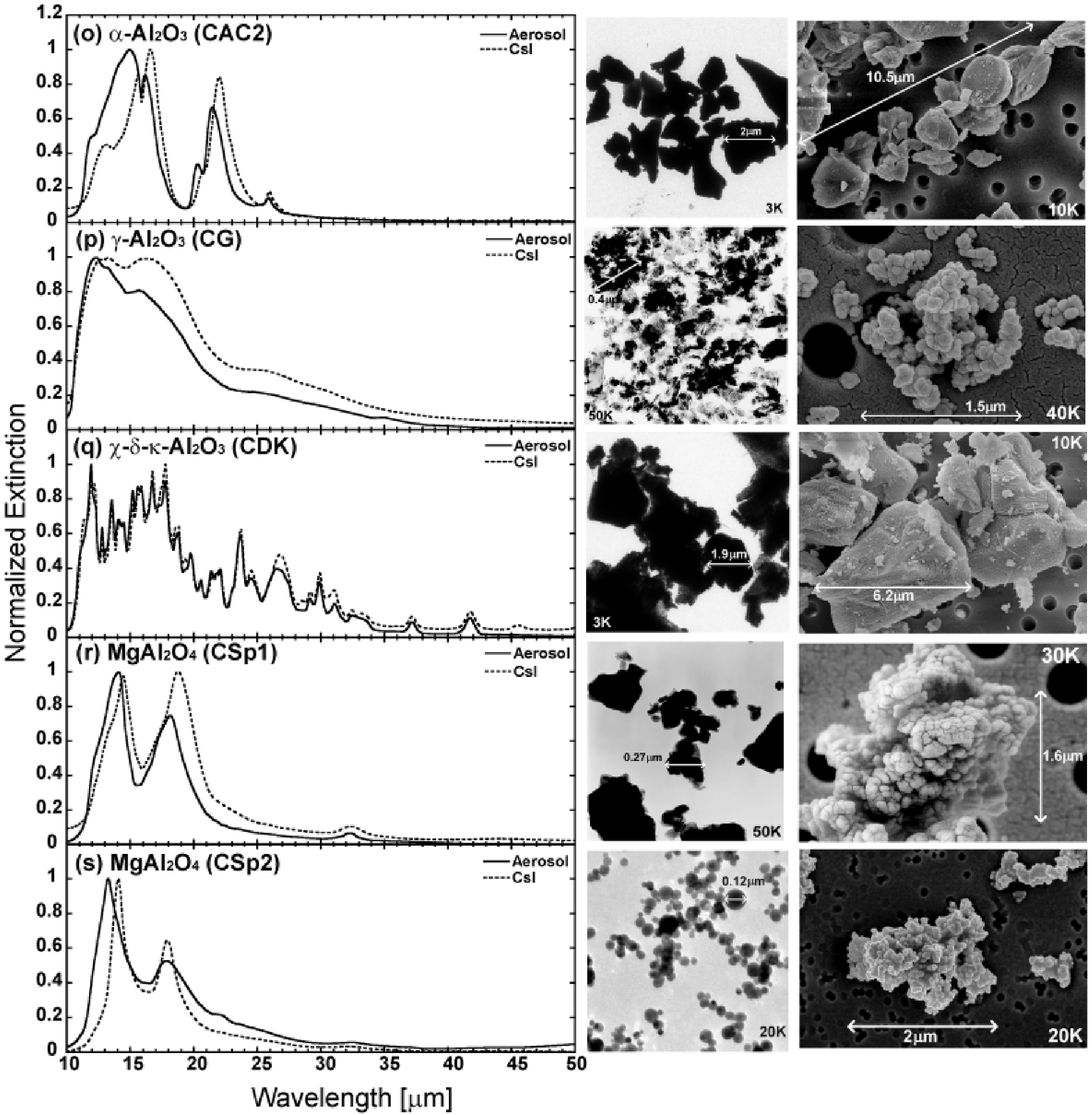}
\addtocounter{figure}{-1}
\caption[\small Normalized extinction vs. wavelength of 19 HTCs]{({\it continued}) Left: Normalized extinction vs. wavelength of (o) $\alpha$-Al$_2$O$_3$ (CAC2); (p) $\gamma$-Al$_2$O$_3$ (CG); (q) $\chi$-$\delta$-$\kappa$-Al$_2$O$_3$ (CCDK); (r) MgAl$_2$O$_4$ (CSp1); (s) MgAl$_2$O$_4$ (CSp2) spectra obtained by aerosol and CsI pellet measurements. Center: TEM images (original powder). Right: SEM images (aerosol particles).}
\end{center}
\end{figure*}
%
\begin{table*}
\small
\caption{The peak positions ($\mu$m) for the aerosol and CsI measurements for all samples except TiO.}
\centering
\footnotesize
\begin{tabular}{c c c c c c c c c c c c c c}
\hline
Chemical formula & Abbrev. & Measurement &   &    &     &  Peaks  & [$\mu$m]  &   &  & & \\
\hline
TiO$_2$     & CR1         & Aerosol  & \bf 13.53 & \bf 19.56 &\bf 24.70 & \bf 30.30 & \bf 35.80 & & & &&\\
            &             & CsI      & 15.61 & 23.77 & 28.13 & 29.81 &       & & & &&\\
\hline        
TiO$_2$     & CR2         & Aerosol  & \bf 15.33 & \bf 23.50 & \bf 27.95 &       &       & & & &&\\
            &             & CsI      & 16.58 & 18.94 & 24.71 & 29.55 &       & & & &&\\
\hline
TiO$_2$     & CR3         & Aerosol  & \bf 16.47 & \bf 19.25 & \bf 23.39 & \bf 28.27 &       & & & &&\\ 
            &             & CsI      & 15.38 & 29.01 &       &       &       & & & &&\\
\hline             
TiO$_2$     & CR4         & Aerosol  & \bf 13.79 & \bf 19.27 & \bf 25.59 & \bf 35.64 &       & & & &&\\   
            &             & CsI      & 15.51 & 19.03 & 24.26 & 29.68 &       & & & &&\\
\hline                        
TiO$_2$     & CA1         & Aerosol  & \bf 13.10 & \bf 16.90 & \bf 28.19 &       &       & & & &&\\             
            &             & CsI      & 15.20 & 19.60 & 28.26 &       &       & & & &&\\
\hline
TiO$_2$     & CA2         & Aerosol  & \bf 13.22 & \bf 16.90 & \bf 27.90 &       &       & & & &&\\
            &             & CsI      & 15.34 & 19.56 & 28.30 &       &       & & & &&\\
\hline
TiO$_2$     & CA3         & Aerosol  & \bf 13.46 & \bf 14.34 & \bf 20.63 & \bf 29.37 &       & & & &&\\
            &             & CsI      & 16.45 & 28.79 &       &       &       & & & &&\\
\hline
TiO$_2$     & CA4         & Aerosol  & \bf 14.58 & \bf 19.89 & \bf 28.26 &       &       & & & &&\\
            &             & CsI      & 17.78 & 29.19 &       &       &       & & & &&\\
\hline
Ti$_2$O$_3$ & Ti$_2$O$_3$ & Aerosol  & \bf 18.22 & \bf 19.90 & \bf 22.62 & \bf 25.54 &       & & & &&\\
            &             & CsI      & 19.28 & 20.30 & 23.11 & 25.72 & 28.53 & &  &  &&\\
\hline
Ti$_3$O$_5$ & Ti$_3$O$_5$ & Aerosol  & \bf 16.21 & \bf 17.91 & \bf 19.73 & \bf 20.36 & \bf 21.69 & \bf 22.07 & \bf 25.44 & \bf 26.06 & & \\
            &             &          & \bf 30.89 & \bf 35.92\\
            &             & CsI      & 16.89 & 18.14 & 19.87 & 20.93 & 24.44 & 25.64 & 26.57 & 31.37 & & \\
            &             &          & 37.14 &       &       &       &       &       &       &       & & \\
\hline
CaTiO$_3$   & CaTiO$_3$   & Aerosol  & \bf 14.85 & \bf 21.67 & \bf 24.56 & \bf 32.13 & \bf 33.71 &       &       &       &       &      \\
            &             & CsI      & 16.89 & 22.18 & 25.95 & 27.86 & 32.30 & 34.16 & 38.96 & 45.70 &       &      \\
\hline
Al$_2$TiO$_5$&Al$_2$TiO$_5$&Aerosol  & \bf 13.48 & \bf 15.48 & \bf 16.31 & \bf 19.96 & \bf 21.33 & \bf 28.38 &       &       &       &      \\
             &             &CsI      & 9.56 & 11.48 & 13.59 & 16.54 & 20.67 & 21.61 & 28.72 &  &       &      \\ \hline
\hline
$\alpha$-Al$_2$O$_3$& CAC1 & Aerosol & \bf 12.65 & \bf 15.37 & \bf 16.16 & \bf 16.91 & \bf 20.14 & \bf 21.57 & \bf 22.48 & \bf 25.92 &       &      \\
                    &      & CsI     & 14.31 & 14.46 & 16.39 & 20.67 & 21.91 & 25.91 &       &       &       &      \\
\hline
$\alpha$-Al$_2$O$_3$& CAC2 & Aerosol & \bf 14.92 & \bf 16.16 & \bf 20.30 & \bf 21.45 & \bf 25.89 &       &       &       &       &      \\
                    &      & CsI     & 13.15 & 15.63 & 16.59 & 20.88 & 22.01 & 25.95 &       &       &       &      \\
\hline                    
$\gamma$-Al$_2$O$_3$& CG   & Aerosol & \bf 12.32 & \bf 13.13 & \bf 15.71 &   &       &       &       &       &       &      \\
                    &      & CsI     & 12.69 & 13.60 & 16.02 & 24.06 &       &       &       &       &       &      \\
\hline
$\chi$-$\delta$-$\kappa$-Al$_2$O$_3$& CDK & Aerosol & \bf 11.93 & \bf 12.79 & \bf 13.56 & \bf 14.08 & \bf 14.47 & \bf 15.19 & \bf 15.87& \bf 16.76 &\\                      &      &          & \bf 17.74 & \bf 18.77 & \bf 19.32 & \bf 19.77 & \bf 20.56 & \bf 21.38 & \bf 22.07 & \bf 23.69 &       &    \\
                    &      &         & \bf 24.48 & \bf 26.62 & \bf 29.15 & \bf 29.89 & \bf 31.09 & \bf 32.46 & \bf 37.20 & \bf 41.70 &       &    \\    
                    &      & CsI     & 11.45 & 11.93 & 12.19 & 12.81 & 13.57 & 14.07 & 14.26 & 14.52 &       &    \\
                    &      &         & 15.21 & 15.62 & 15.88 & 16.76 & 17.78 & 18.58 & 19.29 & 19.78 &       &    \\
                    &      &         & 20.58 & 21.42 & 21.78 & 22.09 & 23.69 & 24.53 & 26.80 & 28.53 &       &    \\
                    &      &         & 29.15 & 29.90 & 31.06 & 32.48 & 33.19 & 37.18 & 41.75 & 45.53 &       &    \\
\hline
\hline
MgAl$_2$O$_4$ & CSp1       & Aerosol & \bf 14.08 & \bf 18.14 & \bf 32.30 &     &    &   &    &    &       \\
              &            & CsI     & 14.45 & 18.80 & 32.34 &       &       &       &       &       &       \\
\hline
MgAl$_2$O$_4$ & CSp2       & Aerosol & \bf 13.26 & \bf 17.79 & \bf 32.61 &     &     &     &     &     &      \\
              &            & CsI     & 14.07 & 17.91 & 32.31 &       &       &       &       &       &        \\
\hline
\end{tabular}
\end{table*}
\indent
The spectrum of CR1 (Fig.~4(b)) shows a similar band profile 
to that of CR4 in Fig.~4(e). 
Although the particle sizes of these samples are different 
from each other, there are two major similarities. 
The individual particle shape is irregular with roundish edges. 
The agglomerate state is composed of many elongated and porous 
agglomerates. 
It is a characteristic of aerosol particles that 
chainlike agglomerates are formed by charged particles \citep{Hi99}.
While the individual particle size of the CR4 sample is smaller 
than that of CR1, the agglomerate size of CR4 is much larger than 
that of CR1. 
As the particle size decreases, it becomes more difficult to remove 
particles from surfaces when the relationship between adhesive and 
separating forces are taken into account \citep{Hi99}. 
Differences in the extinction band profile in the region 13-19~$\mu$m 
may be especially influenced by the agglomerate state.
Increasing the agglomerate size may cause the broadening of the band 
in this region towards longer wavelengths. As a result, a sharp 
fall at 13~$\mu$m cannot be seen in the spectrum of CR4. In other words, 
the secondary peaks at 19~$\mu$m increase markedly.\\
\indent
Comparing CR1 and CR2 (Fig.~4(c)), the individual particle sizes 
are nearly the same, but not the shape. As in the case of forsterite 
(Tamanai et al.~in prep.), irregularly shaped particles (not round edge 
ones) have a tendency to produce a relatively distinctive single peak at 
a somewhat longer wavelength. 
The secondary peak at 19 $\mu$m is not seen 
in the aerosol spectrum of CR2.\\
\indent
Unlike CR1, CR2, and CR4, CR3 (Fig.~4(d)) does not show any 
clear peaks between 10 and 50~$\mu$m in wavelength. 
The secondary peak at 19 $\mu$m seems to dominate. 
When the agglomerate state is close-packed, a broader band 
profile can be produced and hide the peaks from view. 
It is possible to confirm this trend via theoretical calculations 
as well \citep{ta06a}. \\
\indent
The theoretical models in the Rayleigh limit of a continuous 
distribution of ellipsoids (CDE) \citep{bh83} and of spherical 
particles have been utilized to estimate the magnitude of the 
extinction profiles for the rutile samples (CR2 \& CR4).
Two CDE methods have been applied to verify 
the shape effect on spectra. The CDE2 calculates the 
extinction efficiency of the most likely near-spherical 
particle shapes whereas the CDE1 assumes all ellipsoidal
shapes with equal probability \citep[see more details in][]{Os92,Fab01}.\\
\indent
Fig.~5 shows the spectra taken from the aerosol 
measurements (CR2 \& CR4) together with the CDE1, 
CDE2, and Rayleigh calculations. The diameter of the model 
particles is 0.2~$\mu$m and the optical constants from 
\citet{Ri85} are applied for the theoretical calculations.
The spectrum calculated for the spherical particles exhibit 
a sharp-pointed peak with the narrowest bandwidth. As the 
particles become ellipsoidal shapes, the peak undergoes a 
red shift, and the bandwidth gradually broadens (CDE1 \& CDE2). 
These calculations demonstrate the possible shape effect on the measured 
band profiles. The CR2 spectrum in Fig.~5 is between 
the spectra of the CDE1 and CDE2 in the 12-20 $\mu$m wavelength range. 
Fig.~4(c) images show that the CR2 particles have an irregular shape; 
however, the band profiles of the CR2 particles, especially at 23.5 and 
28 $\mu$m are much closer to that calculated for near-spherical shapes. 
Conversely, although the CR4 particles are similarly of near-spherical 
shape (Fig.~4(e)), the CR4 measured band profile does not fit well with 
either the CDE1 or the CDE2 spectrum. 
A nearly trapezoidal profile is seen in the wavelengths 
between 13 and 19~$\mu$m. 
The short wavelength peak corresponds well to spherical particles.
The long wavelength peak is not exactly reproduced by the CDE2. 
The same is true for the profile at longer wavelengths. 
As already mentioned, this behavior may be caused by the 
agglomerate effect because it is more difficult to separate 
particles as particles decrease in size. 
\begin{figure}
\begin{center}
\includegraphics[scale=0.38]{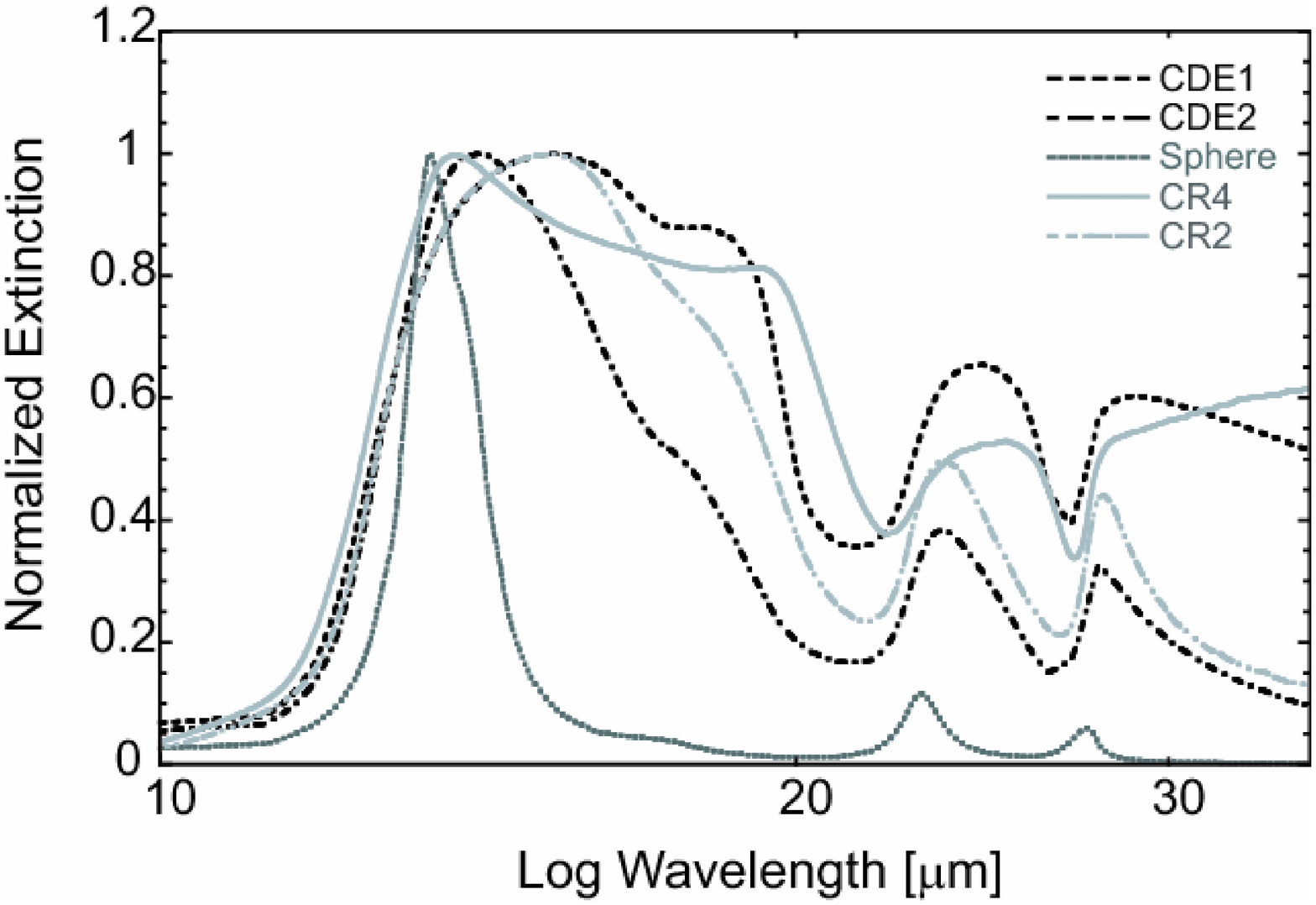}
\caption[\small Normalized extinction vs. wavelength]{Normalized extinction spectra of the two different rutile samples (CR2 \& CR4) obtained by aerosol measurements are compared with three different calculated spectra (CDE \& spheres) in air. (Note: CDE1 for equal weight of all ellipsoidal shapes  \& CDE2 for preferentially near-spherical particles)}
\end{center}
\end{figure}
The effect of agglomeration is also clearly seen in the anatase 
spectra, specifically in the 12-25 $\mu$m band complex.\\
\indent
CA1 (Fig.~4(f)) and CA2 (Fig.~4(g)) have very similar particle conditions. 
They have a particle shape with rounded edges, forming an elongated and 
porous type of agglomerate, and micron size particles. \\
\indent
CA3 (Fig.~4(h)) particles which are nanometer-sized, on the other hand, 
form many close-packed agglomerates which produce a broadening effect 
on the band profile. 
Although the particle size of the CA3 and CA4 (Fig.~4(i)) samples are 
very much alike, the agglomerate state is different. 
Whereas CA3 particles form more close-packed agglomerates, CA4 is
composed of not only close-packed agglomerates, but also a large amount 
of porous agglomerates. 
Consequently, the secondary band at 21 $\mu$m is much stronger for CA3.\\
\indent
The CAC1 aerosol spectrum, in contrast, has a peak at short wavelengths, 
indicating near spherical shape and a trapezoidal profile because of stronger 
agglomeration. \\
\indent
\citet{Beg97} suggested that the 13~$\mu$m 
band profile of $\alpha$-Al$_2$O$_3$ is susceptible 
to particle shape. 
The wavelength range between 13 and 20~$\mu$m of 
$\alpha$-Al$_2$O$_3$ is sensitively affected by 
morphological effects as in the case of rutiles 
\citep{ta07}. 
CAC1 (Fig.~4(n)) and CAC2 (Fig.~4(o)) samples differ in 
both the shapes and sizes of the particles. 
While the CAC1 particles have an irregular shape with rounded 
edges, CAC2 ones have an irregular shape with very sharp edges 
(like crushing a stone into small pieces by a hammer). As in the 
case of rutile (CR1 and CR2) samples, the sharp edged CAC2 
particles give rise to a distinctive single peak at 14.9 $\mu$m 
in the aerosol measurements. \\
\indent
CSp1 (Fig.~4(r)) and CSp2 (Fig.~4(s)) particles form a large number 
of agglomerates with very similar agglomerate states. 
However, CSp2 is composed of near-spherical grains, whereas CSp1 
contains large sharp-edged particles.
It is plausible that the individual 
particle shape exerts a stronger influence on these spectra than the 
agglomerate state and particle sizes. 
In accordance with that expectation, the
spherical shaped particles produce peaks at relatively shorter 
wavelengths and with narrower bandwidth compared to the sharp-edged 
ones (see also Fig.~5). 
The peak for CSp2 particles is located at 13.26~$\mu$m which is 0.82~$\mu$m 
red-shifted compared to the peak of CSp1.\\
\indent
In general, strong extinction peaks are known to originate in geometrical 
resonances of the particles. 
Strong bands rather than the weaker ones are tremendously influenced by 
morphological effects and these effects are seen also in this investigation.
Unfortunately it is virtually impossible to quantify the relative 
importance of each morphological effect. 
A simulation study of shape effects will be published in a forthcoming 
paper (Mutschke et al. in prep.).
\subsection{Quantitative measurements}
We examine the extinction strengths caused by particle 
shapes and embedding media via CDE and Rayleigh calculations. 
The model spectra of forsterite, rutile, and spinel calculated in 
both vacuum and CsI media are shown in Fig.~6. 
All the quantitative spectra are affected by the CsI in the same way.
The spectra in CsI medium exhibit a factor of approximately 1.5 higher extinction strengths than those in vacuum. 
The extinction strengths increase as the particle configuration becomes spherical (CDE1 $\rightarrow$ sphere).
We checked by CDE calculations that apart from a roughly proportional change of the absolute magnitude, spectra of particles in vacuum have the same relative peak strength as in CsI. \\
\begin{figure*}
\begin{center}
\includegraphics[scale=0.58]{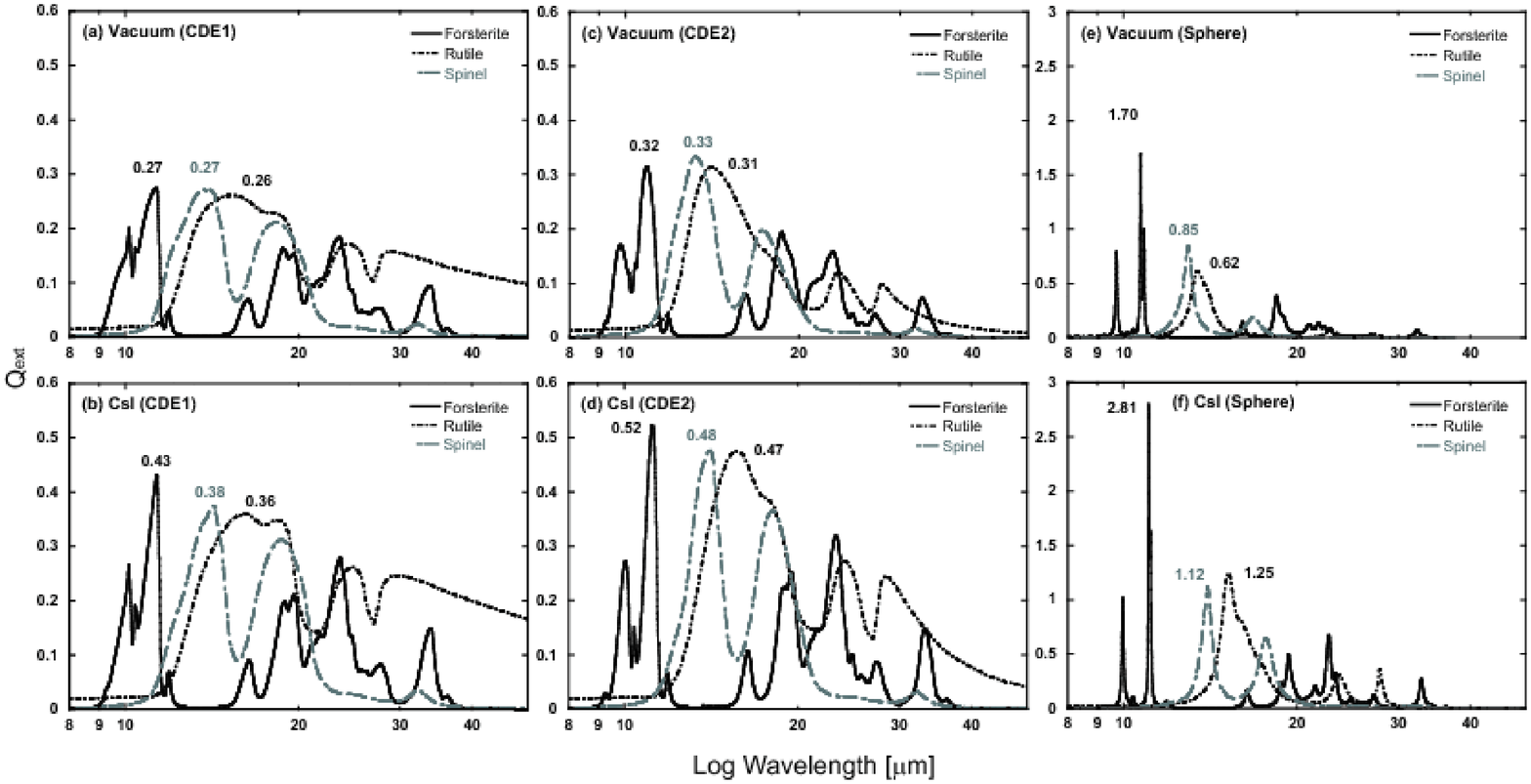}
\caption[\small Extinction efficiency vs. wavelength]{Extinction efficiency vs. wavelength of forsterite, rutile, and spinel spectra in vacuum (upper plots) and CsI medium (bottom plots) obtained by CDE1 (a \& b), CDE2 (c \& d), and sphere (e \& f) calculations (r=0.1~$\mu$m). Numerical extinction strengths at each peak are given in each plot. }
\end{center}
\end{figure*}
\begin{figure}
\begin{center}
\includegraphics[scale=0.38]{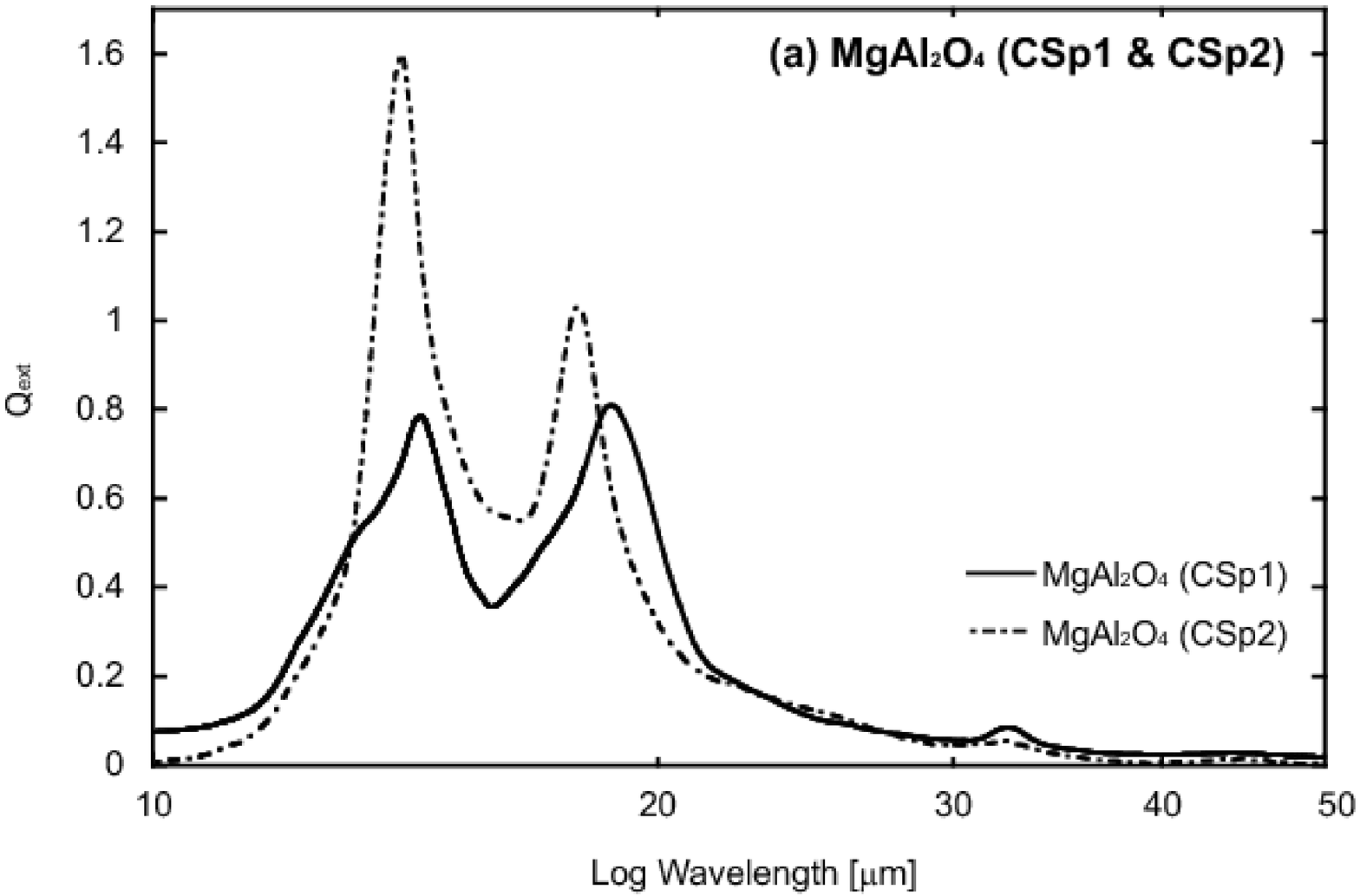}
\includegraphics[scale=0.38]{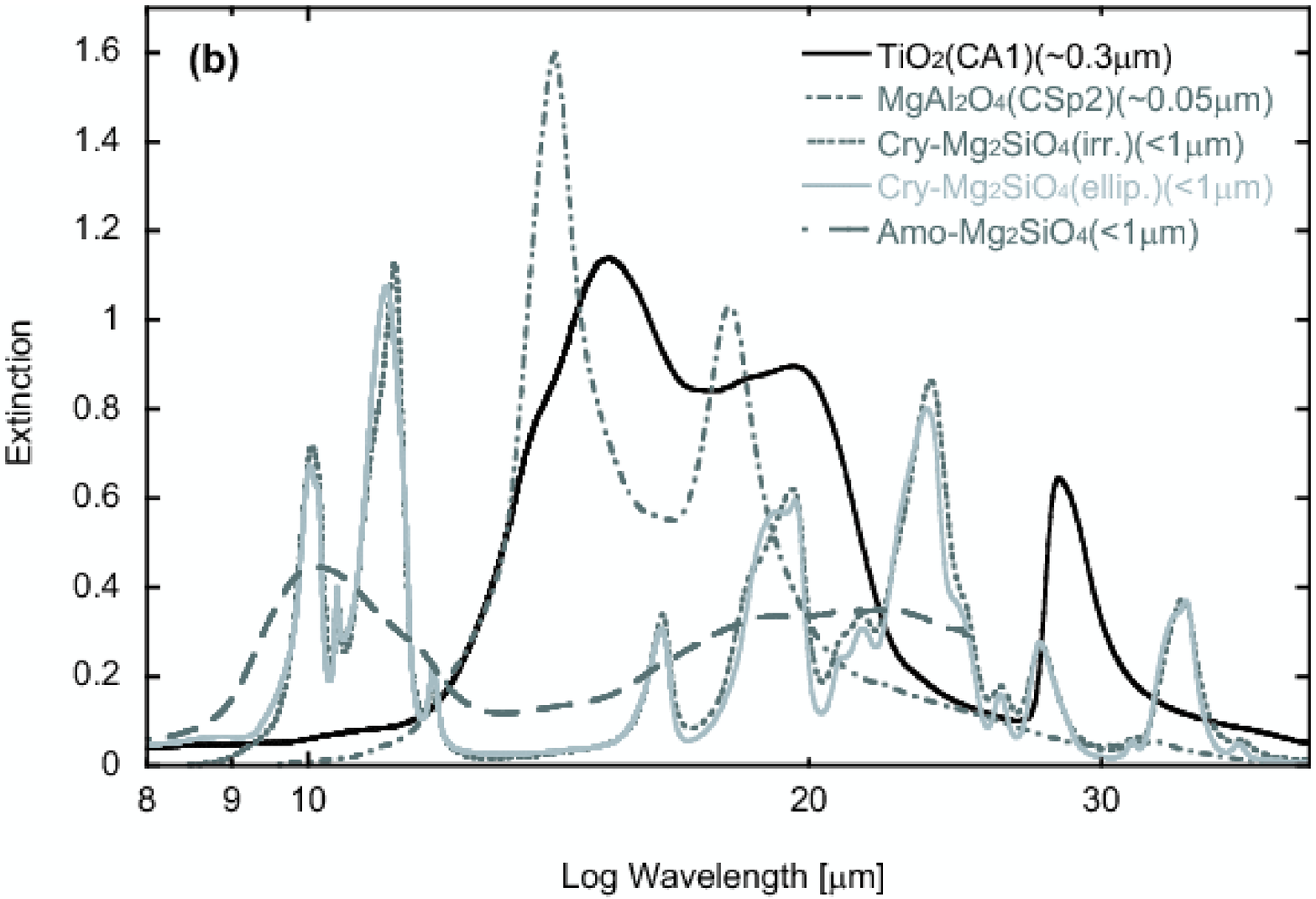}
\caption[\small Extinction vs. wavelength]{(a) Extinction efficiency vs. wavelength of two MgAl$_2$O$_4$ spectra (CSp1 \& CSp2) obtained by CsI pellet measurements. (b) Extinction vs. wavelength of five different spectra (CA1, CSp2, crystalline forsterites (irregular and ellipsoidal shaped particles), and amorphous Mg$_2$SiO$_4$) obtained by CsI pellet measurements.}
\end{center}
\end{figure}
\indent
A comparison between the two MgAl$_2$O$_4$ (CSp1 \& CSp2) samples 
are shown in Fig.~7(a). 
Although identical amounts of samples are embedded in the CsI pellets, 
these spectra show an enormous difference of the peak strength in the spectrum. 
The extinction at 14~$\mu$m for CSp2 is almost twice as much as that of CSp1. 
It is probable that the difference in extinction between the CSp1 and CSp2 samples 
is chiefly caused by morphological effects, especially shape.\\
\indent
We compare the quantitative measured spectrum of amorphous and 
crystalline Mg$_2$SiO$_4$ (ellipsoidal and irregular shaped forsterites) 
with TiO$_2$ (CA1) and MgAl$_2$O$_4$ (CSp2) (Fig.~7(b)).
In the legend of Fig.~7(b), the particle sizes are listed as well. 
These HTC spectra are of much stronger extinction strength than the 
amorphous Mg$_2$SiO$_4$ spectrum. \\
\indent
Note also that the extinction strengths of forsterite differ from 
the theoretical calculations for the CsI measured spectrum (Fig.~7(b)).
As the forsterite particle shape becomes spherical, the strength of 
the 11~$\mu$m peak becomes stronger and more prominent than the others 
in the calculations, but the strength of spinel in the measured spectrum 
is in fact the strongest in this comparison. 
The peak position of measured ellipsoidal forsterite takes place at 
11.14~$\mu$m which is closer to the peak obtained by the calculation for 
spherical particles (11.06~$\mu$m) compared to that of irregular shaped 
forsterite (11.27~$\mu$m). 
The bandwidth of measured ellipsoidal forsterite is much broader than 
that of the calculated band profile, and the measured forsterite band 
profile does not contain any prominent peaks. 
On the other hand, the peak position of measured spinel is located at 
14.07~$\mu$m whereas the calculated peak lies at 13.99~$\mu$m 
($\Delta$$\lambda$=0.08~$\mu$m). 
The bandwidth of calculated spinel is slightly narrower than the 
measured one. 
However, the band profiles of the measured and calculated spinel 
spectra are comparable. 
This result directly affects the extinction strength. 
The ellipsoidal forsterite particles are not spherical enough to 
produce the stronger extinction strength compared to the spherical 
spinel. 
The greater extinction strength seen for forsterite is due to the 
shape effect of individual particles upon the extinction strength.\\
\indent
According to \citet{Mi04}, amorphous silicates are conceivably 
the dominant species of grains, with a small amount of crystalline 
silicates in oxygen-rich AGB stars. 
As shown in Fig.~7(b), amorphous Mg$_2$SiO$_4$ produces an ineffective extinction feature. 
If the amount of crystalline Mg$_2$SiO$_4$ is much less than that of oxides, 
it may be possible to detect some pronounced features from oxide dust grains 
in mid-IR region even if amorphous Mg$_2$SiO$_4$ is abundant.
\section{Astrophysical implication}
We selected three AGB objects which exhibit strong 13, 19.5, and 28~$\mu$m features. 
ISO emission spectra of S~Pav, Y~UMa, and g~Her are shown in Fig.~8 together with 
the spectra obtained by the aerosol measurements of anatase (CA1 \& CA4), spinel 
(CSp2), tialite (Al$_2$TiO$_5$) and amorphous Mg$_2$SiO$_4$ \citep{ta06b} in the 
wavelength range between 8 to 45~$\mu$m. 
Basic properties and detailed observed data analysis information of these AGB 
objects are given in \citet{Po99}, \citet{Po02}, and \citet{Fab01}. \\
\indent
The emission spectra of S Pav and Y UMa (Fig.~8(a) \& (b)) show a strong and broad 
9-40~~$\mu$m background. 
An 11~$\mu$m peak, which is clearly seen in both spectra, may originate from 
amorphous Al$_2$O$_3$ \citep{Ca02}. 
In contrast, the emission spectrum of g-Her (Fig.~8(c)) has a weaker background 
and a clear 10~$\mu$m amorphous silicate band \citep{Mi04}. 
The 10 $\mu$m band appears as a shoulder only in the spectra of S Pav and Y UMa. 
These spectra exhibit bands at 13, 16-22 (equally strong, with substructures), 
28, and 32~$\mu$m (weak, especially in S Pav; apparently quite strong in g-Her). 
The relative strength of the bands may be a temperature effect. \\
\begin{figure}
\begin{center}
\includegraphics[scale=0.66]{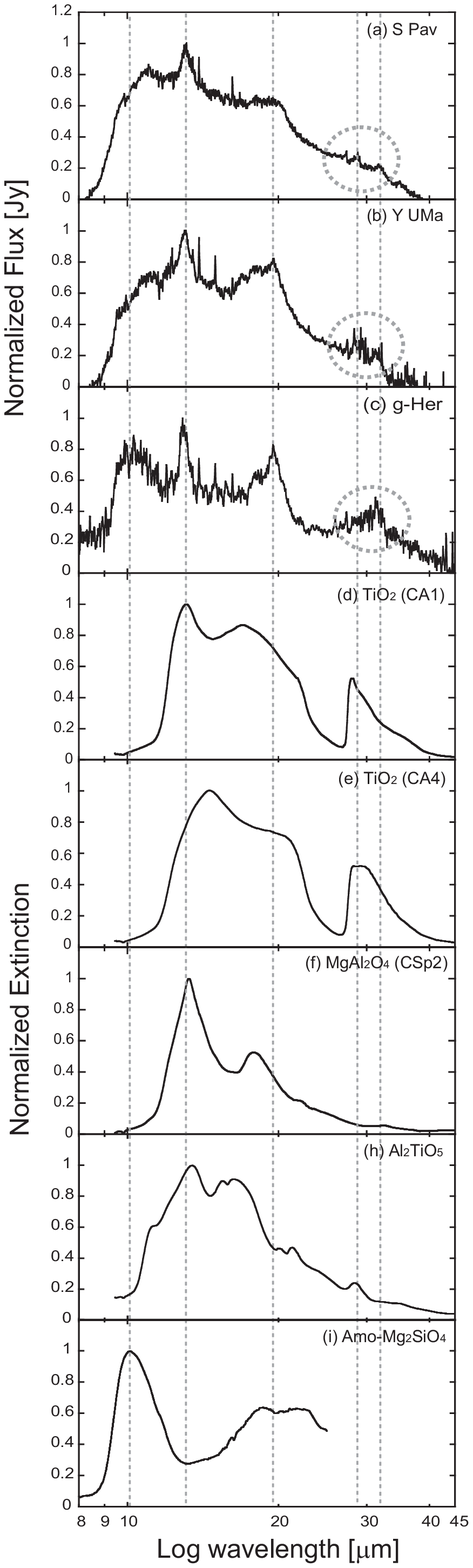}
\caption[\small Comparison between observed and experimental data]{Comparison of the observed emission features for the three AGB stars with the experimental results. The normalized emission spectrum of (a)~S~Pav; (b)~Y~UMa; (c)~g~Her are compared with the aerosol measurements of (d)~anatase (CA1); (e)~anatase (CA4); (f)~spinel (CSp2); (g)~tialite (Al$_2$TiO$_5$); (h)~amorphous Mg$_2$SiO$_4$. The gray dotted lines denote 11, 13, 19.5, 28 and 32~$\mu$m wavelengths.}
\end{center}
\end{figure}
\indent
A strong 13~$\mu$m feature appears in all three objects as well as 
19.5 and 28~$\mu$m features. 
Anatase, spinel and tialite produce peaks 
around 13~$\mu$m, but the anatase (CA1) peak (13.1~$\mu$m) corresponds 
best with the 13~$\mu$m feature of the observed spectra. 
However, the 13~$\mu$m feature in the observed spectra is rather a 
prominent peak. 
It is possible that spinel (like CSp2) enhances the 
protuberance of the 13~$\mu$m peak. \\
\indent
As mentioned in Sec.~3.2, the wavelength range between 13 and 20~$\mu$m of the  
TiO$_2$ spectra is finely influenced by morphological effects. 
Fig.~8(d) and (e) show the two types of anatase spectra that exhibit 
different band profiles caused by morphological effects (Table~1). 
As the morphological state of the particles varies, the band 
profile is noticeably changed as well. 
Consequently, it may be that 
anatase particles with a different morphological state are able to enhance 
the 13~$\mu$m peak and/or to produce a clearer peak around 19~$\mu$m. \\
\indent
Additionally, we consider that tialite plays an important 
role in the observed spectra.
When all these HTCs (Al$_2$O$_3$, TiO$_2$, MgAl$_2$O$_4$) are 
present in the same environment, tialite and more complex 
titanate dust grain formation is expected (see sec.~A.6.).
According to \citet{Bus96}, MgAl$_2$O$_4$ and 
TiO$_2$ react into Mg$_{0.5}$AlTi$_{1.5}$O$_5$ 
particles which are formed at a temperature above $\approx$1423~K. 
These behave like nuclei 
for further titanate (e.g. AlTiO$_3$, MgTiO$_3$, MgTi$_2$O$_5$, Al$_2$TiO$_5$) 
formation and growth. 
Consequently, MgAl$_2$O$_4$ produces an effect on the mechanism of Al$_2$TiO$_5$ formation. 
If spinel dust grains exist in an environment, Mg containing titanate may form \citep[see also][]{Po03}.
This is also an interesting point to associate with a subsequent 
condensation process in the outflow of AGB stars. 
From the spectroscopic point of view, although the 13~$\mu$m peak of tialite is not 
as prominent of the peak in CSp2, this peak is expected to enhance 
the strength of the emission in the observed spectra.\\
\indent
As to the peak at 19.5~$\mu$m, none of the measured spectra produced a peak  
adequate to explain the observed spectra. 
However, the observed peak at 19.5~$\mu$m 
might be contributed by anatase dust grains because the three observed spectra 
in Fig.~8 have a common feature, namely that the emission strength of the 13~$\mu$m feature 
correlates with that of the 19.5~$\mu$m peak.
g-Her (Fig.~8(c)) shows the narrowest and sharpest 13 and 19~$\mu$m peaks 
among the observed spectra while S Pav exhibits a  
broader peak at 13~$\mu$m, and then the 19.5~$\mu$m peak becomes more like a small shoulder. 
\citet{Slo03} discussed about the correlated dust features at 13, 20 and 28~$\mu$m and 
concluded that the 20 and 28~$\mu$m features originated in silicates. 
\citet{Po02} proposed that the 19.5~$\mu$m feature carrier would be 
Mg$_{0.1}$Fe$_{0.9}$O in relation to the band profiles (the band width, shape and position), 
but the formation of Mg$_{0.1}$Fe$_{0.9}$O in the circumstellar environment 
is not yet well understood.
We expect that all three peaks (13, 19.5 and 28~$\mu$m) come from a single source. 
The observed spectra undergo a slight change in the band profiles (e.g. peak shift, 
band broadening, and the strength of extinction feature) by other relatively less 
influential sources.
In order to clarify the source of the 19.5~$\mu$m feature, it is necessary to take 
morphological effects into account and to carry out further investigation.\\
\indent
Another point to be investigated is the band feature beyond 19.5~$\mu$m.
The observed spectra exhibit a gradual fall beyond 19.5~$\mu$m, and excessive 
extinction features cannot be seen in this wavelength region. 
In fact, rutile is the most stable material out of the three crystalline 
phases of TiO$_2$ (see sec.~A.2.). 
However, the anatase spectra fit better 
with these observed spectra since they do not show any peaks around 25~$\mu$m,  
unlike the rutile spectra (Fig.~4(b)-(e)). 
The minor peaks circled ($\approx$~28~$\mu$m) in Fig.~8(a-c) fit well with 
the anatase features (CA1 \& CA2) as well as with tialite.\\
\indent
Furthermore, as mentioned above, the 10~$\mu$m feature originates 
from amorphous silicate \citep{Mi04}. 
In this investigation, the amorphous Mg$_2$SiO$_4$ spectrum obtained by 
aerosol measurement compares well with the 10~$\mu$m feature, especially 
for the spectrum of g-Her.

\section{Summary and discussions}
Spectroscopic extinction measurements of HTCs are performed 
in the wavelength range between 10 and 50~$\mu$m by making 
use of the aerosol technique in order to obtain the band 
profiles without any medium effect. 
We focus particularly on morphological effects on the measured 
spectra to obtain a new perspective on the 13, 19.5, and 28~$\mu$m 
carriers in the spectra of AGB stars. 
We summarize the main results of this paper as follows:
\begin{itemize}
\item There is a tendency for peaks to appear at longer wavelengths 
due to the influence of the medium on the surface modes.
The effect is expected \citep{bh83}, but is very 
strong for many HTCs.
The maximum peak position difference between the aerosol and 
CsI pellet measurement is 3.20~$\mu$m in wavelength in the case 
of the TiO$_2$ anatase (CA4) sample.
\item Particles which have spherical or roundish shapes produce 
larger differences between the spectra measured by the aerosol and 
the CsI pellet techniques as compared to the irregular shaped 
particles.
\item Roundish grains tend to produce double peaked (rectangular) profiles 
 in comparison with irregular shaped grains. 
\item The agglomerate state may strongly influence the strength of the strong bands. 
The extinction band profile between 13 and 19.5~$\mu$m undergoes a substantial change 
for different agglomerate states (rutile \& anatase).
\item Although band profiles are influenced by the morphological 
effects, it is practically unfeasible to quantify 
the relative importance of each morphological effect.
\item In comparison to TiO$_2$, 
Al$_2$O$_3$ creates a larger disparity in the band profiles. 
In particular, three different crystal structures have been detected by 
a X-ray diffraction analysis in $\chi$-$\delta$-$\kappa$-Al$_2$O$_3$ that 
produce 25 discernible peaks up to 45~$\mu$m.
\item The absolute strength of extinction strongly depends on  
chemical composition and particle morphology. 
Spherical shaped nano-sized spinel shows the most effective extinction 
in all samples investigated here. 
\item Through the comparison of the observed emission in AGB spectra and the aerosol spectra, 
we find that primarily the roundish edged anatase (CA1) is anticipated to 
contribute to the 13 and 28~$\mu$m peaks. 
Spherical shaped nano-sized spinel (CSp2) plays a role in intensifying the 
13~$\mu$m peak as well as tialite (Al$_2$TiO$_5$).
The carrier for the peak at 19.5~$\mu$m could not be identified well in this investigation. However,
as a possibility, the 19.5~$\mu$m peak might be produced by the same source 
(anatase) when the particle morphology and temperature are taken into account.
\end{itemize} 
In fact, titanium is a significantly less abundant metal by a factor of 
400, compared to Si \citep{Ca73}. 
\citet{Po03} noted that rutile and 
anatase in the 13~$\mu$m region are about 40 times more 
effective absorbers than amorphous silicate, and they predicted that the most 
prominent TiO$_2$ emission will be about 1/10th as strong as the silicate 
dust emission. 
Likewise, our CsI pellet measurements exhibit very similar results, and show that  
anatase radiates approximately 8.5 times more efficiently than amorphous 
Mg$_2$SiO$_4$. 
However, there still a huge difference between the abundance proportions 
of Ti and Si. If the solar abundance can be applied in the 
AGB star environments, the Ti abundance remains unclear.\\
\indent
The abundance of possible condensates has been calculated under the assumption of chemical equilibrium. 
Chemical equilibrium cannot apply to  
the dust condensation process in circumstellar dust shells since 
grain formation takes place in a non-equilibrium kinetic process. 
The non-equilibrium calculations of \citet{FG01} showed that a different mixture of minerals form compared to an equilibrium calculation. 
Although the equilibrium assumption may be very convenient as an approximation to 
understand possible condensates present in certain conditions, 
the detailed condensation process can only be explored with non-equilibrium calculations. 
More dependable non-equilibrium condensation process results are required to study 
further dust grain formation in the circumstellar outflow of AGB stars. \\
\indent
While having a precise knowledge of the absolute strength of extinction 
for each substance is essential to identify the dust grains present in 
dust shells, it is also very important to understand heterogeneous 
dust grain properties. 
We have concentrated only on homogeneous dust grain investigations here. 
Each sample has unique properties which determine the agglomerate state. 
The agglomeration of non-spherical particles increases in comparison to 
spherical particles with the same volume because the non-spherical particles 
have a larger surface area which leads to more collisions by Brownian 
motion \citep[e.g.][]{Hi99,Blu00,KB04,PD06}. 
As the particle shape irregularity becomes much more complicated, the agglomeration 
effect will be more significant. According to a theoretical calculation by \citet{Ze66}, 
ellipsoidal particles have a 35~\% higher agglomeration coefficient than spherical ones 
with the same volume. Hence, heterogeneous grains make both the thermal and kinematic 
agglomeration processes more complicated. An experimental heterogeneous dust 
grain investigation is indispensable to address this problem. \\
\indent
By way of another approach, we will attempt to mix 
crystalline and amorphous samples together so as to 
observe how these mixed particles interact with each 
other and produce an effect on the band profiles. 
We focused only on crystalline dust particles in 
this investigation, since we presume that the 13~$\mu$m 
peak mainly originates from a crystalline species. 
Unlike crystalline materials, amorphous particles tend to 
produce broader bands due to the range of bond lengths 
and angles of atoms in the amorphous structure \citep{Ja03}. 
However, as the three observed AGB spectra (Fig.~8) show 
broad peaks between 10 and 13~$\mu$m, it is recognizable 
that amorphous materials (e.g. silicates) take part in 
the spectra. 
Posch et al. (2002) proposed that amorphous Al$_2$O$_3$ 
contributes a broad "continuum emission" (g Her, Y UMa, 
V1943 Sgr) in the wavelength range between 11 and 15~$\mu$m.
Thus, it is interesting to examine how the strong 
crystalline features are weakened by amorphous materials 
by changing the quantity of each material. \\
\indent
During aerosol measurements, the large and compact agglomerates 
start to be deposited in the cell; therefore, only the remaining 
small and fluffy agglomerates can keep flying in the cell for a long time. 
These aerosol spectra reflect the difference in bandwidth. 
A spectrum from a sample which contains more large agglomerates 
shows broader bands. 
However, the variation of these spectra is not very strong. 
Since the particle densities are too low (10$^6$~particles/cm$^3$) 
and the gas pressure is too high, these experiments are not a 
suitable environment to observe the growth of particles by agglomeration.\\
\indent
Concerning the quantitative aspects of the aerosol measurements, 
it is extremely difficult to obtain the exact mass concentration in 
the cell during the measurements. 
One of the possible ways to solve this problem is to derive the mass 
through a theoretical analysis. 
The extinction efficiency of small-sized particles can be calculated by means 
of the form-factor distribution \citep{min06} which can then be fitted 
to the measured spectrum. 
A forthcoming paper will include more discussion of these concepts 
(Mutschke et al. in prep.).\\
\indent
Identification of dust grains in astrophysical environments is not straightforward. 
The band profiles are strongly influenced by physical and chemical factors of the dust 
grains which interact with each other and make the identification all the more complicated. 
It is essential to carry out condensation experiments and more 
detailed non-equilibrium condensation theoretical approaches for the circumstellar 
dust shells around AGB stars. \\
\indent
The infrared extinction spectra database can be found at {\it http://elbe.astro.uni-jean.de}. 
The database includes the numerical 
extinction efficiency data obtained by both the aerosol and CsI pellet 
measurements, SEM and TEM images, basic properties, plots, and peak positions 
of each sample. 

\begin{acknowledgements}
We express to the members of the pathology department at FSU Jena, Dr. S. Nietzsche, F. Steiniger, and C. Kamnitz our gratitude for the technical support of the TEM and SEM. We are thankful to Dr. D.R. Alexander at ISU and Dr. D. Hilditch at FSU Jena (TPI) for proofreading and insightful opinions and comments.
We are grateful to G. Born for her assistance with sample preparation and W. Teuschel for his technical support of our experimental devices. Our project has been supported by Deutsche Forschungsgemeinschaft (DFG) under the grant MU 1164/6. 
\end{acknowledgements}

\appendix
\section{Physical properties of oxides}
\subsection{TiO}
Titanium oxide, TiO$_{\rm_x}$ 
(0.7$\leq$x$\leq$1.25), forms in a cubic crystal structure \citep{Jun84}. 
TiO is the one-to-one composition of the TiO$_x$ series and has a defective 
NaCl crystal structure which is composed of identical numbers of random 
vacancies in the cation and anion sites in cubic symmetry at temperatures 
higher than 1273~K \citep{Fu89,we91}. However, the cubic 
crystal structure of TiO$_x$ can be transformed into various ordered 
structures such as cubic, tetragonal, orthorhombic, and monoclinic lattices.
The transformation depends strongly on oxygen content and temperature 
conditions \citep{Jun84,Gus91}. The melting point of TiO is 
2023~K \citep{Li90} which is the lowest temperature among the Ti-compounds investigated here.\\
\indent
Solid TiO forms from TiO$_2$ and H$_2$ gases through the chemical reaction in the circumstellar outflow of M stars
\begin{eqnarray*}
   \mbox{TiO$_2$} \mbox{(gas)} + \mbox{H$_2$} &\longrightarrow& \mbox{TiO} \mbox{(solid)}  + \mbox{H$_2$O} 
\end{eqnarray*}
\citep{GS98}.
\subsection{TiO$_2$}
Titanium dioxide (titanium(IV) oxide or titania) (TiO$_2$) occurs in three different crystal structures, which are stable at atmospheric pressure. They are rutile, anatase, and brookite. 
Although the anatase and brookite phases are stable at lower 
temperature than rutile, both transform to rutile at approximately 
1188~K (anatase) and 1023~K (brookite) \citep{BB92}. 
Rutile is the most stable form of the three phases.
It belongs to the tetragonal crystal system and has 
a melting point at 2123~K \citep{Li90}. 
Anatase is a member of the tetragonal crystal 
system as well. The density of anatase is approximately 10~\% 
less than that of rutile \citep{Lin95}. 
The melting point of anatase, which is about 2108~K \citep{Re97}, 
is slightly lower than that of rutile. \\
\indent
In the crystal structure system, rutile and anatase consist of the basic structural unit of the TiO$_6$ octahedron. Each Ti$^{+4}$ ion occupies the center of the oxygen octahedra, so that it is surrounded by 6 O$^{2-}$ ions. Conversely, since each O$^{2-}$ ion is enclosed by 3 Ti$^{4+}$ ions, the Ti:O ratio will be 3:6. In both cases, TiO$_2$ octahedra are distorted, and anatase has more distortions than rutile \citep[see details in e.g.][]{Mo95,KH93,Mu04}.\\
\indent
TiO$_2$ solid would form from the gas phase via
\begin{eqnarray*}
   \mbox{TiO$_2$} \mbox{(gas)} &\longrightarrow& \mbox{TiO$_2$} \mbox{(solid)} 
\end{eqnarray*}
\citep{GS98}. Solid TiO$_2$ (rutile) contributes to Al$_2$TiO$_5$ formation (see~2.2.6).\\
\indent
In addition, when a deposited TiO thin film underwent annealing 
in air, anatase and rutile structures have been shown at a temperature 
of 793~K \citep{Zri08}, which indicates possible TiO$_2$ 
formation from TiO via an annealing process. \\
\indent
The optical properties of TiO$_2$ (rutile, brookite, anatase) are found in 
\citet{Po03}.
\subsection{Ti$_2$O$_3$}
Dititanium trioxide (Ti$_2$O$_3$) has a trigonal crystal structure. 
O$^{2-}$ ions are positioned in a hexagonal close-packed structure and 
Ti$^{3+}$ ions occupy 2/3 of the octahedral interstices \citep{we91}. 
Ti$_2$O$_3$ has a corundum-type structure (see sec.~A.7.) at all 
temperatures \citep{SE62}. The melting point is 2503~K,  
which is much higher than that of TiO$_2$. \\
\indent
A possible solid Ti$_2$O$_3$ formation pathway from the gas phase is 
\begin{eqnarray*}
   \mbox{2TiO$_2$} \mbox{(gas)} + \mbox{H$_2$} &\longrightarrow& \mbox{Ti$_2$O$_3$} \mbox{(solid)} + \mbox{H$_2$O}
\end{eqnarray*}
\citep{GS98}.
Reflectance measurements of a single Ti$_2$O$_3$ crystal sample at 
room-temperature have been performed by \citet{Luc77}. 
The measurements were made for the two primary polarizations, namely 
with the crystalline c-axis perpendicular and parallel to the plane of incidence. 
The TO-phonon frequencies obtained with the perpendicular measurements 
were 19.6, 22.2, 26.6, and 35.7~$\mu$m and 18.6, 19.9, 25.6, 
and 35.6~$\mu$m for the LO modes. Likewise, the parallel 
measurements of the TO-phonon frequencies were shown at 22.3 and 29.2~$\mu$m, 
and 18.1 and 28.5~$\mu$m for the LO modes \citep[see also][]{Po03}. 
\subsection{Ti$_3$O$_5$}
Trititanium pentoxide is one of the Ti$_n$O$_{2n-1}$ (n$\geq$4) series.
\citet{AM59} ascertained the Ti$_3$O$_5$ 
crystal structure at room temperature to be monoclinic. The crystal 
structure of Ti$_3$O$_5$ is composed of a three dimensional array of TiO$_6$ 
octahedra sharing edges and vertices \citep{we91}. However, Ti$_3$O$_5$ is 
also a substance which undergoes phase transitions of crystal structures 
during heating. A transition from monoclinic to orthorhombic phases has been  
seen at a temperature of 514~K \citep{Ono98}. \\
\indent
Ti$_3$O$_5$ may form via the chemical reaction
\begin{eqnarray*}
   \mbox{3TiO$_2$} \mbox{(gas)} + \mbox{H$_2$} &\longrightarrow& \mbox{Ti$_3$O$_5$} \mbox{(solid)} + \mbox{H$_2$O}
\end{eqnarray*}
\citep{GS98}. The melting point of Ti$_3$O$_5$ is 2033~K 
\citep{In99} which is slightly lower than 
anatase and rutile. 
The stretching vibration of TiO$_6$ octahedra takes 
place in the wavelength range between 4.3 and 16.7~$\mu$m 
(600-700~cm$^{-1}$) \citep{Gil93} similarly to other 
Ti-compounds.
\subsection{CaTiO$_3$}
According to the relative abundances of dust grain calculations (sec.~2.3) (Fig.~2), CaTiO$_3$ 
(perovskite) appears at approximately 1700~K in chemical equilibrium 
before a large quantity of silicates starts to condense out of the gas. 
Solid perovskite may form via the chemical reaction 
\begin{eqnarray*}
   \mbox{TiO$_2$} \mbox{(gas)} + \mbox{Ca} + \mbox{H$_2$O} &\longrightarrow& \mbox{CaTiO$_3$} \mbox{(solid)} + \mbox{H$_2$}.
\end{eqnarray*}
Ca exists as a free atom in the gas phase because Ca atoms are not 
able to form high bond energy molecules \citep{GS98}. 
The melting point of perovskite is approximately 2248~K (Alfa Aesar Catalog) 
which is the second highest value among Ti-compounds.\\
\indent
Perovskite has an orthorhombic crystal system and a face centered 
cubic lattice (fcc) \citep[e.g.][]{RC00}. Eight smaller Ti$^{+4}$ 
cations are located at each corner of the cubic structure, and the large 
Ca$^{+2}$ cation occupies the center. Twelve O$^{2-}$ anions are 
positioned between each Ti$^{4+}$ cation. Thus, the Ca$^{2+}$ cation 
is surrounded by the 12 O$^{2-}$ and 8 Ti$^{4-}$ cations with a 
coordinate distance to form the cubic close-packed structure which 
is sustained only if the 12-fold coordinated Ca$^{2+}$ cation is 
larger than oxygen. Otherwise, the cubic crystal structure undergoes 
a distortion to transform into other crystal forms such as octahedra, 
tetragonal, orthorhombic, and monoclinic forms depending on the temperature. 
Once the face-centered cubic lattice loses its shape due to a change in 
temperature down to the Curie point, perovskite becomes a ferroelectric 
material that has a very high dielectric constant \citep[e.g.][]{DH06}.
The origin of CaTiO$_3$ IR absorption mainly arises from Ti-O stretching 
modes around 18.2~$\mu$m (550~cm$^{-1}$) and 22.5~$\mu$m (445~cm$^{-1}$), 
Ti-O-Ti bending mode at 55.6~$\mu$m (180~cm$^{-1}$), and the cation-TiO$_3$ 
lattice mode at 66.7~$\mu$m (150~cm$^{-1}$) (Perry \& Khanna~1964). \\
\indent
\citet{Po03} derived optical constants of a natural 
CaTiO$_3$ crystal from reflectance measurements.
\subsection{Al$_2$TiO$_5$}
We considered that aluminum titanate (tialite or Al$_2$TiO$_5$) is a 
very interesting species to investigate, though it has not been discovered 
yet in any astronomical objects. Al$_2$TiO$_5$ can be formed above its 
equilibrium formation temperature, 1553~K \citep{FM87,FM88}. 
The formation of Al$_2$TiO$_5$ via the endothermic reaction would be
\begin{eqnarray*}
   \mbox{$\alpha$-Al$_2$O$_3$} \mbox{(corundum)} + \mbox{TiO$_2$} \mbox{(rutile)} &\longrightarrow& \mbox{$\beta$-Al$_2$TiO$_5$} 
\end{eqnarray*}
\citep[e.g.][]{Wo88,Ji03}. 
The melting point of Al$_2$TiO$_5$ is rather high (2133~K), as 
for Al$_2$O$_3$ and TiO$_2$ \citep{Ji03}. Although it is possible 
to maintain Al$_2$TiO$_5$ under metastable conditions at room temperature, 
the decomposition takes place at a temperature above 1123~K \citep{Wo88}. 
The reaction between corundum and rutile into Al$_2$TiO$_5$ begins at 1473~K. 
As the temperature increases to 1523~K, Al$_2$TiO$_5$ dominates 
in quantity \citep{Ji03}. 
Al$_2$TiO$_5$ belongs to the orthorhombic pseudobrookite-type 
crystal structure, which is composed of MO$_6$ octahedra sharing 
edges and vertices (M denotes metal) \citep{we91}. 
Ti-O and Al-O vibrations induce absorption at wavelengths between 
13.3 and 25~$\mu$m (750-400~cm$^{-1}$). In detail, the stretching 
modes from octahedral AlO$_6$ appear between 13.3 and 16.7~$\mu$m 
(750-600~cm$^{-1}$), and the bending mode occurs approximately at 
22.2~$\mu$m (450~cm$^{-1}$). By the same token, TiO stretching 
vibration modes are induced below 13.7~$\mu$m (730~cm$^{-1}$) in 
wavelength and tetrahedral coordinated Al-O produces peaks around 
11.8 to 13.3~$\mu$m (850-750~cm$^{-1}$) \citep[e.g.][]{PT71,St04}.\\
\indent
While Al$_2$TiO$_5$ may be sufficiently less abundant to not be visible in 
observed spectra, the formation of Al$_2$TiO$_5$ is possible 
via the chemical reaction of corundum and rutile. 
\subsection{Al$_2$O$_3$}
Aluminum oxides crystallize into the form of corundum ($\alpha$-Al$_2$O$_3$), 
which is the second hardest natural mineral. This hardness might be caused 
by the strong and short O-Al ionic bonds which attract O$^{2-}$ ions and 
Al$^{3+}$ ions close together in order to form a tremendously hard and 
dense close-packed crystal structure. $\alpha$-Al$_2$O$_3$ is formed 
from aluminum hydroxides via many phase transitions by increasing calcination temperature. 
Take boehmite (AlO(OH)), for instance. 
The dehydration reaction pathway of boehmite is that it is first transformed into 
$\gamma$-Al$_2$O$_3$ (750~K) $\rightarrow$ $\delta$-Al$_2$O$_3$ (1050~K) 
$\rightarrow$ $\theta$-Al$_2$O$_3$ ($\approx$ 1200~K) 
$\rightarrow$ $\alpha$-Al$_2$O$_3$ ($\approx$ 1300~K) \citep{WM87}. 
$\delta$-Al$_2$O$_3$ has a tetragonal crystal structure like $\gamma$-Al$_2$O$_3$. 
On the one hand, in the dehydration reaction pathway of gibbsite (Al(OH)$_3$) is first changed to 
$\chi$-Al$_2$O$_3$ (473-1173~K) $\rightarrow$ $\kappa$-Al$_2$O$_3$ (1173-1273~K)
$\rightarrow$ $\alpha$-Al$_2$O$_3$ ($>$ 1273~K) \citep{Co07}. 
$\kappa$-Al$_2$O$_3$ has hexagonal and orthorhombic crystal structures 
whereas the crystal systems of $\chi$-Al$_2$O$_3$ are hexagonal 
and cubic \citep{Bha04}. 
All the phases are metastable polymorphs of transition Al$_2$O$_3$ 
except $\alpha$-Al$_2$O$_3$ which is always the end product. 
A reverse transition is also able to produce one or more metastable 
or transition Al$_2$O$_3$ molecules. $\gamma$-Al$_2$O$_3$ is the most common 
resultant from $\alpha$-Al$_2$O$_3$ by 
cooling \citep{Sa00}. \\
\indent
$\alpha$-Al$_2$O$_3$ has a rhombohedral (or trigonal) crystal structure and is considered to be one of the early condensation species in a cooling gas environment of solar composition \citep{Beg97}. 
The melting point is in the range between 2273 and 2303~K \citep{Li90}. In the crystal structure of $\alpha$-Al$_2$O$_3$, the O$^{2-}$ ions are arranged in a hexagonal close-packing structure, and the Al$^{3+}$ ions occupy 2/3 of the octahedral interstices. One Al$^{3+}$ ion is surrounded by 6 O$^{2-}$ ions whereas one O$^{2-}$ ion is enclosed within 4 Al$^{3+}$ ions. Thus, the Al:O ratio will be 4:6 \citep[e.g.][]{Mu04}. \\
\indent
$\gamma$-Al$_2$O$_3$ has a tetragonal crystal structure \citep{Sa58}. $\gamma$-Al$_2$O$_3$ is a so-called "defect-spinel" because it has basically the same structure as spinel \citep[e.g.][]{SM99}. Since $\gamma$-Al$_2$O$_3$ does not have Mg cations, Al cations occupy the Mg sites. As a result, the stoichiometry becomes Al$_3$O$_4$. It is necessary to remove 8/3 Al$^{3+}$ ions from the 24 ions available in the spinel unit cell in order to correct the  stoichiometry. Then, the unit cell contains 96 O$^{2-}$ ions, 64 Al$^{3+}$ ions, and 8 spinel cation vacancies. These vacancies are randomly distributed; exact locations have not been verified \citep[e.g.][]{SM99}.\\
\indent
The equilibrium between a condensed phase and vapor phase can be represented in the case of $\alpha$-Al$_2$O$_3$ as
\begin{eqnarray*}
   \mbox{2Al} \mbox{(gas)} + \mbox{3O} \mbox{(gas)} &\longrightarrow& \mbox{$\alpha$-Al$_2$O$_3$}\mbox{(solid)} 
\end{eqnarray*}
\citep{GL74}.\\
\indent
Al-O vibrations are induced in the mid-IR region \citep{Beg97}. 
LO modes were confirmed around 15.74, 17.57, 22.62, and 25.97~$\mu$m 
(635, 569, 442, and 385~cm$^{-1}$) with the electric field polarized 
perpendicular to the c-axis (ordinary ray) whereas LO modes around 11.11, 
16.0, 20.83, and 25.77~$\mu$m (900, 625, 480, and 388~cm$^{-1}$) were 
observed with the ordinary ray. Likewise, the TO mode peaks at approximately 
15.29, 17.15, and 25.00~$\mu$m (654, 583, and 400~cm$^{-1}$) appeared 
with the electric field polarized parallel to the c-axis (extraordinary ray).
The LO modes of the extraordinary ray could be detected at about 11.48 and 19.53~$\mu$m 
(871 and 512~cm$^{-1}$) \citep[see details in][]{Ba63}. 
Especially for $\gamma$-Al$_2$O$_3$, there exist vibrational frequencies in 
the range from 12.5 to 14.29~$\mu$m (680-500~cm$^{-1}$) and from 14.71 to 
20~$\mu$m (800-700~cm$^{-1}$) \citep{Sa95}. \\
\indent
Optical properties of various $\alpha$-Al$_2$O$_3$ samples have been analyzed 
by many groups \citep[e.g.][]{Ba63,Lo73,Qu85,Ge91} as well as 
$\gamma$-Al$_2$O$_3$ \citep[e.g.][]{Ch88,Ko95}, $\delta$- and $\theta$-Al$_2$O$_3$ \citep{Ku05}, 
and amorphous Al$_2$O$_3$ \citep[e.g.][]{Ch88,Beg97}.
\subsection{MgAl$_2$O$_4$}
MgAl$_2$O$_4$ (spinel) is known to be an important ferromagnetic material, 
which has very high permeability \citep[e.g.][]{DH06} and a cubic close-packed anion arrangement \citep[e.g.][]{Ro01} like most metallic crystals. The general formula of the spinel group can be described as XY$_2$O$_4$. X would be replaced with Mg$^{2+}$, Fe$^{2+}$, Ti$^{4+}$, Zn$^{2+}$, or Mn$^{2+}$. 
Y may be occupied by Al$^{3+}$, Fe$^{3+}$, Fe$^{2+}$, or Cr$^{3+}$. \\
\indent
In the case of magnesium aluminum spinel (MgAl$_2$O$_4$), 
the oxygens form a face-centered cubic closed-packed array 
along planes in the structure. Then, Mg$^{2+}$ ions are 
placed in tetrahedral interstices (1/8 occupied) whereas 
Al$^{3+}$ ions sit in octahedral (1/2 occupied) sites in 
the lattice. Eight tetrahedral sites are occupied by Mg$^{2+}$ 
ions, and 16 octahedral sites are filled by Al$^{3+}$ ions 
surrounded by 32 O$^{2-}$ ions in a unit cell of the spinel. 
In consequence, spinel has 56 ions per unit cell
\citep[see details in][]{KH93,Fab01}.\\
\indent
Spinel can be found as a natural mineral and can also be syntheszed. 
Mg$^{2+}$ and Al$^{3+}$ ions are originally well ordered 
in natural spinel lattices; however, the spinel crystal undergoes 
disordering by annealing. Disordering of these 
cations is seen in both annealed and synthetic crystal spinels.
The structural transition occurs at temperatures between 1023 and 
1073~K due to the disordering 
\citep[see details in][]{Sch72,Fab01}. 
The melting point of spinel is 2408~K \citep{Li90}.\\
\indent
Possible spinel formation pathways depend on temperature. At higher temperature, spinel grains may be formed via the reaction
\begin{eqnarray*}
   \mbox{Mg} \mbox{(gas)} + \mbox{2Al} \mbox{(gas)} + \mbox{4O} \mbox{(gas)} &\longrightarrow& \mbox{MgAl$_2$O$_4$} \mbox{(solid)}. 
\end{eqnarray*}
As the temperature drops around 1758~K, Al is not able to remain in the gas phase together with three oxygens for every two atoms. Thus, it is conceivable that spinel forms via a chemical reaction with corundum at lower temperature ($\approx$1500~K)
\begin{eqnarray*}
   \mbox{Mg} \mbox{(gas)} + \mbox{O} \mbox{(gas)} + \mbox{Al$_2$O$_3$} \mbox{(cry)} &\longrightarrow& \mbox{MgAl$_2$O$_4$} \mbox{(cry)} 
\end{eqnarray*}
("cry" denotes crystalline)\citep{GL74}.\\
\indent
In addition, spinel formation has been confirmed in shock induced experiments by making use of single crystals of  corundum and periclase (MgO) \citep{PA94}. They reported that a MgAl$_2$O$_4$ layer is formed at the boundary between these two crystals under particular conditions. Hence, it is also possible that spinel forms via a solid-solid reaction with some shock impacts
\begin{eqnarray*}
   \mbox{MgO} \mbox{(cry)} + \mbox{Al$_2$O$_3$} \mbox{(cry)} &\longrightarrow& \mbox{MgAl$_2$O$_4$} \mbox{(cry)}.
\end{eqnarray*}
\indent
The fundamental (one photon) lattice vibrations of spinel generate bands between 12.5 and 33.3~$\mu$m (800-300~cm$^{-1}$) \citep{TT91}. \\
\indent
Optical properties of MgAl$_2$O$_4$ have been investigated 
by e.g. \citet{TT91}, \citet{Chi00}, and \citet{Fab01}. 
\citet{Chi00} and \citet{Fab01} derived the 
optical constants of both natural and synthetic spinels from 
reflectance measurements.

\begin{thebibliography}{}
\bibitem[Allard et al.(2001)]{Al01}
Allard, F., Hauschildt, P.H., Alexander, D.R., Tamanai, A., \& Schweitzer A. 2001, \apj, 556, 357
\bibitem[${\rm\AA}$sbrink \& Magn\'{e}li(1959)]{AM59}
${\rm\AA}$sbrink, S., \& Magn\'{e}li, A. 1959, Acta Cryst., 12, 575
\bibitem[Barker(1963)]{Ba63}
Barker, A.S.Jr. 1963, Phys.Rev., 132, 1474
\bibitem[Begemann et al.(1997)]{Beg97}
Begemann, B., Dorschner, J., Henning, Th., \& Mutschke, H. 1997, \apj, 476, 199
\bibitem[Bhattacharya et al.(2004)]{Bha04}
Bhattacharya, I.N., Das, S.C., Mukherjee, P.S., Paul, S., \& Mitra, P.K. 2004, Scand.J.Metall., 33, 211
\bibitem[Blum et al.(2000)]{Blu00}
Blum, J., Wurm, G., Poppe, T., et al. 2000, Phys.Rev.Lett., 85, 2426
\bibitem[Brandes \& Brook(1992)]{BB92}
Brandes, E.A., \& Brook, G.B. 1992, Smithells Metal Reference Book (Butterworth-Heinemann Ltd., Oxford)
\bibitem[Bohren \& Huffman(1983)]{bh83}
Bohren, C.F., \& Huffman, D.R. 1983, Absorption and Scattering of Light by Small Particles (John Wiley \& Sons Inc., N.Y.)
\bibitem[Buscaglia et al.(1996)]{Bus96}
Buscaglia, V., Delfrate, M.A., Leoni, M., \& Bottino, C. 1996, J.Mater.Sci, 31, 1715
\bibitem[Cameron(1973)]{Ca73}
Cameron, A.G.W. 1973, \ssr, 15, 121 
\bibitem[Cami(2002)]{Ca02}
Cami, J. 2002, Ph.D. thesis, University of Amsterdam, The Netherlands
\bibitem[Chihara et al.(2000)]{Chi00}
Chihara, H., Tsuchiyama, A., Koike, C., \& Sogawa, H. 2000, in ASP Conference Series Disks, Planetesimals and Planets, eds. F. Garzon, C. Eiroa, D. de Winter, \& T.J. Mahoney, Vol.219, 150
\bibitem[Chihara et al.(2002)]{Chi02}
Chihara, H., Koike, C., Tsuchiyama, A., Tachibana, S., \& Sakamoto, D 2002, \aap, 391,267
\bibitem[Chu et al.(1988)]{Ch88}
Chu, Y.T., Bates, J.B., White, C.W., \& Farlow, G.C. 1988, J.Appl.Phys., 64, 3727
\bibitem[Cl\'{e}ment et al.(2003)]{Cl03}
Cl\'{e}ment, D., Mutschke, H., Klein, R., \& Henning, Th. 2003, \apj, 594, 642
\bibitem[Coelho et al.(2007)]{Co07}
Coelho, A.C.V., Santos, H.de S., Kiyohara, P.K., Marcos, K.N.P., \& Santos, P.de S. 2007, Matar.Res., 10, 183
\bibitem[DePew et al.(2006)]{De06}
DePew, K., Speck, A., \& Dijkstra, C. 2006, \apj, 640, 971 
\bibitem[Drain(1988)]{dr88}
Draine, B.T. 1988, \apj, 333, 848
\bibitem[Dorschner et al.(1978)]{do78}
Dorschner, J., Friedemann, C., \& G\"{u}rtler, J. 1978, Astron.Nachr., 299, 6
\bibitem[Douglas \& Ho(2006)]{DH06}
Douglas, B.E., \& Ho, S.M. 2006, Structure and Chemistry of Crystalline Solids (Springer, N.Y.)
\bibitem[Fabian et al.(2001)]{Fab01}
Fabian, D., Posch, Th.,  Mutschke, H., Kerschbaum, F., \& Dorschner 2001, \aap, 373, 1125
\bibitem[Ferguson et al.(2005)]{Feg05}
Ferguson, J.W. et al. 2005, \apj, 623, 585 
\bibitem[Ferrarotti \& Gail(2001)]{FG01}
Ferrarotti, A.S., \& Gail, H.-P. 2001, \aap, 371, 133 
\bibitem[Freundenberg \& Mocellin(1987)]{FM87}
Freundenberg, B., \& Mocellin, A. 1987, J.Am.Ceram.Soc., 70, 33
\bibitem[Freundenberg \& Mocellin(1988)]{FM88}
Freundenberg, B., \& Mocellin, A. 1988, J.Am.Ceram.Soc., 71, 22
\bibitem[Fujimura et al.(1989)]{Fu89}
Fujimura, T. et al. 1989, High Pressure Research, 1, 213
\bibitem[Gail \& Sedlmayr(1998)]{GS98}
Gail, H.P., \& Sedlmayr, E. 1998, Faraday Discuss., 109, 303
\bibitem[Gail \& Sedlmayr(1999)]{GS99}
Gail, H.P., \& Sedlmayr, E. 1999, \aap, 347, 594
\bibitem[Gervais(1991)]{Ge91}
Gervais, F. 1991, Handbook of Optical Constants of Solids II, ed. Palik, E.D. (Academic Press Inc., Orland) 1035
\bibitem[Gillet(1993)]{Gil93}
Gillet, P., Guyot, F., Price, G.D., Tournerie, B., \& Le Cleach, A. 1993, Phys. Chem. Minerals, 20, 1993
\bibitem[Gusev(1991)]{Gus91}
Gusev, A.I. 1991, phys.stat.sol~(b), 163, 17
\bibitem [Glaccum(1995)]{Gla95}
Glaccum, W 1995, in ASP Conf. Proc. 73, Proc. Airborne Astronomy Haas, ed. J.A. Davidson, \& E.F.Erickson (ASP, San Francisco), 395
\bibitem[Greshake et al.(1996)]{Gre96}
Greshake, A., Bischoff, A., \& Putnis, A. 1996, Lunar \& Planetary Science, 27, 463
\bibitem[Greshake et al.(1998)]{Gre98}
Greshake, A., Bischoff, A., \& Putnis, A. 1998, Meteoritics \& Planetary Science, 33, 75
\bibitem[Grossman \& Larimer(1974)]{GL74}
Grossman, L., \& Larimer, J.W., 1974, Rev. of Geophys.\& Space Phys., 12, 71
\bibitem[Habing \& Olofsson(2004)]{HO04}
Habing, H.J., \& Olofsson, H. 2004, Asymptotic Giant Branch Stars, ed. H.J. Habing, \& H. Olofsson (Springer-Verlag, N.Y.), 1
\bibitem[Henning \& Mutschke(2000)]{HM00}
Henning, Th., \& Mutschke, H. 2000, in Thermal Emission Spectroscopy and Analysis of Dust, Disks, and Regoliths, eds.~M.L.~Sitko, A.L.~Sprague, and D.K.~Lynch, 196, 253
\bibitem[Hinds(1999)]{Hi99}
Hinds, W.C. 1999, Aerosol Technology: Properties, Behavior, and Measurement of Airborne Particles, (John Wiley \& Sons Inc., N.Y.)
\bibitem[International Advanced Materials(1999)]{In99}
International Advanced Materials 1999, http://www.iamaterials.com/compounds
/titaniumoxide4.htm
\bibitem[J\"{a}ger et al.(1994)]{Ja94}
J\"{a}ger, C., Mutschke, H., Begemann, B., Dorschner, J., \& Henning, Th. 1994, \aap, 292, 641
\bibitem[J\"{a}ger et al.(2003)]{Ja03}
J\"{a}ger, C., Dorschner, J., Mutschke, H., Posch, Th., \& Henning, Th. 2003, \aap, 408, 193
\bibitem[Jeong et al.(1999)]{Je99}
Jeong, K.S., Winters, J.M., \& Sedlmayr, E. 1999, in IAU symposium Asymptotic Giant Branch Stars, eds. T.~LeBerte, A.~Lebre, \& C.~Waelkens, Nr.191, 233
\bibitem[Jianu et al.(2003)]{Ji03}
Jianu, A., Stanciu, L., Groza, J.R., Lathe, Ch., \& Burkel, E., 2003, Nucl.Instrum.Methods B, 199, 44
\bibitem[Jun et al.(1984)]{Jun84}
Jun, S.G., Bursill, L.A., Yoshida, K., Yamada, Y., \& Ota, H. 1984, Acta Cryst., B40, 549
\bibitem[Klein \& Hurlbut(1993)]{KH93}
Klein, C., \& Hurlbut, C.S.Jr. 1993, Manual of Mineralogy, (John Wiley \& Sons, Inc., N.Y.)
\bibitem[Koike et al.(1981)]{Ko81}
Koike, C., Hasegawa, H., Asada, N., \& Hattori, T. 1981, Ap\&SS, 79, 77
\bibitem[Koike \& Hasegawa(1987)]{KH87}
Koike, C., \& Hasegawa, H. 1987, Ap\&SS, 134, 361
\bibitem[Koike et al.(1995)]{Ko95}
Koike, C. et al. 1995, \icarus, 114, 203
\bibitem[Kozasa \& Sogawa(1997)]{KS97}
Kozasa, T., \& Sogawa, H. 1997, Ap\&SS, 255, 437
\bibitem[Krause \& Blum(2004)]{KB04}
Krause, M., \& Blum, J. 2004, Phys.Rev.Lett., 93, 021103
\bibitem[Kurumada et al.(2005)]{Ku05}
Kurumada, M., Koike, C., \& Kaito, C. 2005, \mnras, 359, 643
\bibitem[Lide(1990)]{Li90}
Lide, D.R. 1990, CRC Handbook of Chemistry and Physics (CRC Press Inc., Boston)
\bibitem[Linsebigler et al.(1995)]{Lin95}
Linsebigler, A.L., Lu, G., \& Yated, J.T.Jr. 1995, Chem. Rev., 95, 735
\bibitem[Loewenstein et al.(1973)]{Lo73}
Loewenstein, E.V., Smith, D.R., \& Morgan, R.L. 1973, Applied Optics, 12, 398
\bibitem[Lucovsky et al.(1977)]{Luc77}
Lucovsky, G., Sladek, R.J., \& Allen, J.W. 1977, Phys. Rev. B, 16, 5452
\bibitem[Mackowski \& Mishchenko(1996)]{mm96}
Mackowski, D.W., \& Mishchenko, M.I. 1996, J.Opt.Soc.Am.A, 13, 2266
\bibitem[Mie(1908)]{mie08}
Mie, G. 1908, Ann.Phys., 25, 377
\bibitem[Millar(2004)]{Mi04}
Millar, T.J. 2004, Asymptotic Giant Branch Stars, ed. H.J. Habing, \& H. Olofsson (Springer-Verlag, N.Y.), 247
\bibitem[Min et al.(2005)]{min05}
Min, M., Hovenier, J.W., de Koter, A., Waters, L.B.F.M., \& Dominik, C. 2005, \icarus, 179, 158
\bibitem[Min et al.(2006)]{min06}
Min, M., Hovenier, J.W., Dominik, C., de Koter, A., \& Yurkin, M.A. 2006, \jqsrt, 97, 161
\bibitem[Mishchenko(1990)]{mi90}
Mishchenko, M.I. 1990, Ap\&SS, 164, 1
\bibitem[Mo et al.(1995)]{Mo95}
Mo, S.-D, \& Ching, W.Y. 1995, Phy.Rev.B, 51, 13023
\bibitem[Muraishi(2004)]{Mu04}
Muraishi, H. 2004, Basic Solid State Chemistry for Inorganic Materials (Sankyo Shuppan Inc., Tokyo)
\bibitem[Nittler et al.(2005)]{Nit05}
Nittler, L.R., Alexander, C.M.O'D., Stadermann, F.J., \& Zinner, E.K. 2005, in 36$^{th}$ Annual Lunar and Planetary Science Conference, Abst.~No.2200
\bibitem[Nittler et al.(2008)]{Nit08}
Nittler, L.R., et al. 2008, \apj, 686, 1524
\bibitem[Onaka et al.(1989)]{Ona89}
Onaka, T., de Jong, T., \& Willems, F.J. 1989, \aap, 218, 169
\bibitem[Onoda(1998)]{Ono98}
Onoda, M. 1998, J. Solid State Chem., 136, 67
\bibitem[Orofino et al.(1991)]{Oro91}
Orofino, V., et al. 1991, \aap, 252, 315
\bibitem[Ossenkopf et al.(1992)]{Os92}
Ossenkopf, V., Henning, Th., \& Mathis, J.S. 1992, \aap, 261, 567
\bibitem[Papoular et al.(1998)]{Pa98}
Papoular, R., Cauchetier, M., Begin, S., \& LeCaer, G. 1998, \aap, 329, 1035
\bibitem[Paszun \& Dominik(2006)]{PD06}
Paszun, D., \& Dominik, C. 2006, \icarus, 182, 274
\bibitem[Perry \& Khanna(1964)]{PK64}
Perry, C.H., \& Khanna, B.N. 1964, Phys.Rev.A, 135, 408 
\bibitem[Posch et al.(1999)]{Po99}
Posch, Th., et al. 1999, \aap, 352, 609
\bibitem[Posch et al.(2002)]{Po02}
Posch, Th., Kerschbaum, F., Mutschke, H., Dorschner, J., \& J\"{a}ger, C. 2002, \aap, 393, L7
\bibitem[Posch et al.(2003)]{Po03}
Posch, Th. et al. 2003, \apjs, 149, 437
\bibitem[Potter \& Ahrens(1994)]{PA94}
Potter, D.K., \& Ahrens, T.J. 1994, Geophys.Res.Lett., 21, 721
\bibitem[Preudhomme \& Tarte(1971)]{PT71}
Preudhomme, J., \& Tarte, P. 1971, Spectrochimica Acta, 27A, 961
\bibitem[Purcell \& Pennypacker(1973)]{pp73}
Purcell, E.M., \& Pennypacker, C.R. 1973, \apj, 186, 705
\bibitem[Querry(1985)]{Qu85}
Querry, M.R. 1985, Optical Constants (Chemical Research and Development Center, Maryland)
\bibitem [Reade Advanced Materials(1997)]{Re97}
Reade Advanced Materials 1997, http://www.reade.com/Products/Minerals$\_$and
$\_$Ores/anatase.html
\bibitem[Redfern \& Carpenter(2000)]{RC00}
Redfern, S.A.T., \& Carpenter, M.A. 2000, Transformation Processes in Minerals (Mineralogical Society of America, Washington DC)
\bibitem[Ribarsky(1985)]{Ri85}
Ribarsky, M.W. 1985, Handbook of Optical Constants of Solids, ed. Palik, E.D. (Academic Press Inc., Orland), 795
\bibitem[Rohrer(2001)]{Ro01}
Rohrer, G.S. 2001, Structure and bonding in crystalline materials (Cambridge Univ. Press, Cambridge)
\bibitem[Saalfeld(1958)]{Sa58}
Saalfeld, H. 1958, Clay Min.Bull., 3, 249
\bibitem[Saniger(1995)]{Sa95}
Saniger, J.M. 1995, Materials Letters, 22, 109
\bibitem[Santos et al.(2000)]{Sa00}
Santos, P.S., Santos, H.S., \& Toledo, S.P. 2000, Mater.Res., 3, 104
\bibitem[Schmocker et al.(1972)]{Sch72}
Schmocker, U., Boesch, H.R., \& Waldner, F. 1972, Phys. Lett., 40A, 237
\bibitem[Sedlmayr(1994)]{sed94}
Sedlmayr, E. 1994, in Molecules in the Stellar Environment, ed. U.G.~Jorgensen, 428, 163 (Springer-Verlag, Berlin)
\bibitem[Sharp \& Huebner(1990)]{SH90}
Sharp, C.M., \& Huebner, W.F. 1990, \apjs, 72, 417
\bibitem[Sloan et al.(2003)]{Slo03}
Sloan, G.C., Kraemer, K.E., Goebel, J.H., \& Price, S.D. 2003, \apj, 594, 483 
\bibitem[Speck et al.(2000)]{Sp00}
Speck, A.K., Barlow, M.J., Sylvester, R.J., \& Hofmeister, A.M. 2000, \aaps, 146, 437
\bibitem[Stanciu et al.(2004)]{St04}
Stanciu, L.A., Groza, J.R., Jitianu, A., \& Zaharescu, M. 2004, Mater.Manuf.Process, 19, 641
\bibitem[Straumanis \& Ejima(1962)]{SE62}
Straumanis, M.E., \& Ejima, T. 1962, Acta Cryst., 15, 404
\bibitem[Streitz \& Mintmire(1999)]{SM99}
Streitz, F.H, \& Mintmire, J.W. 1999, Phys.Rev.B, 60, 773
\bibitem[Stroud et al.(2004)]{Str04}
Stroud, R.M, Nittler, L.R., \& Alexander, C.M.O'D. 2004, Sci, 305, 1455
\bibitem[Tamanai et al.(2006a)]{ta06a}
Tamanai, A., Mutschke, H., Blum, J., Neuh\"auser, R. 2006a, \jqsrt, 100, 373
\bibitem[Tamanai et al.(2006b)]{ta06b}
Tamanai, A., Mutschke, H., Blum, J., \& Meeus, G. 2006b, \apj, 648, L147
\bibitem[Tamanai(2007)]{ta07}
Tamanai, A. 2007, Ph.D. thesis, FSU Jena, Germany
\bibitem[Tropf \& Thomas(1991)]{TT91}
Tropf, W.J., \& Thomas, M.E. 1991, Handbook of Optical Constants of Solids II, ed. Palik, E.D. (Academic Press Inc., Orland), 883
\bibitem[Wefers \& Misra(1987)]{WM87}
Wefers, K., \& Misra, C. 1987, Oxides and Hydroxides of Aluminum (Alcoa Research Laboratories, USA)
\bibitem[Wells(1991)]{we91}
Wells, A.F. 1991, Structural Inorganic Chemistry (Clarendon Press, Oxford)
\bibitem[Woignier et al.(1988)]{Wo88}
Woignier, T., Lespade, P., Phalippou, J., \& Rogier R. 1988, J.Non-Cyst.Solids, 100, 325
\bibitem[Zebel(1966)]{Ze66}
Zebel, G. 1966, Coagulation of Aerosols, in Aerosol Science, ed. C.N. Davis, (Academic Press, N.Y.), 31
\bibitem[Zribi et al.(2008)]{Zri08}
Zribi, M., Kanzari, M., \& Rezig, B. 2008, Thin Solid Films, 516, 1476
\end{thebibliography}
\end{document}